\documentclass[journal,twoside]{IEEEtran}
\usepackage{amsfonts,bm,booktabs,cite,graphicx,hyperref,multirow,textcomp,url}
\usepackage{algorithm}
\usepackage{subfigure}
\usepackage{algpseudocode}
\usepackage{amsmath}
\usepackage{mathrsfs}
\usepackage{float}
\usepackage{color}
\usepackage{booktabs}
\usepackage{bbding}
\usepackage[flushleft]{threeparttable}

\makeatletter 
\renewcommand{\@thesubfigure}{\hskip\subfiglabelskip}
 \makeatother
\IEEEpubid{\begin{minipage}{\textwidth}\ \centering
		Copyright \copyright 2025 IEEE.  Personal use of this material is permitted.  Permission from IEEE must be obtained for all other uses, in any current or future media, including reprinting/republishing this material for advertising or promotional purposes, creating new collective works, for resale or redistribution to servers or lists, or reuse of any copyrighted component of this work in other works.
\end{minipage}}

\begin{document}
\title{Prediction and Reference Quality Adaptation for Learned Video Compression}
\author{
	Xihua Sheng, 
	Li Li, \IEEEmembership{Member, IEEE},
	Dong Liu, \IEEEmembership{Senior Member, IEEE},
	Houqiang Li, \IEEEmembership{Fellow, IEEE}\\
\thanks{
Date of current version \today.\par 
This work was supported in part by the Natural Science Foundation of China under Grants 62171429 and 62021001. It was also supported by the GPU cluster built by MCC Lab of Information Science and Technology Institution, USTC.\par
X. Sheng, L. Li, D. Liu, and H. Li are with the MoE Key Laboratory of Brain-inspired Intelligent Perception and Cognition, University of Science and Technology of China, Hefei 230027, China (e-mail: xhsheng@mail.ustc.edu.cn, lil1@ustc.edu.cn, dongeliu@ustc.edu.cn, lihq@ustc.edu.cn). Corresponding author: Li Li.\par
}
}

\markboth{IEEE TRANSACTIONS ON IMAGE PROCESSING}{Sheng \MakeLowercase{\textit{et al.}}: Prediction and Reference Quality Adaptation for Learned Video Compression}

\maketitle
\begin{abstract}
Temporal prediction is one of the most important technologies for video compression. Various prediction coding modes are designed in traditional video codecs. Traditional video codecs will adaptively to decide the optimal coding mode according to the prediction quality and reference quality. Recently, learned video codecs have made great progress. However, they did not effectively address the problem of prediction and reference quality adaptation, which limits the effective utilization of temporal prediction and reduction of reconstruction error propagation. Therefore, in this paper, we first propose a confidence-based prediction quality adaptation (PQA) module to provide explicit discrimination for the spatial and channel-wise prediction quality difference. With this module, the prediction with low quality will be suppressed and that with high quality will be enhanced. The codec can adaptively decide which spatial or channel location of predictions to use. Then, we further propose a reference quality adaptation (RQA) module and an associated repeat-long training strategy to provide dynamic spatially variant filters for diverse reference qualities. With these filters, our codec can adapt to different reference qualities, making it easier to achieve the target reconstruction quality and reduce the reconstruction error propagation. Experimental results verify that our proposed modules can effectively help our codec achieve a higher compression performance.
\end{abstract}
\IEEEpeerreviewmaketitle

\section{Introduction}
With the rapid growth of various emerging video applications, such as internet protocol television (IPTV), live streaming, and online meetings, video data has contributed to most of the internet traffic. The large amount of video data brings large transmission and storage costs. Therefore, it puts forward high requirements for compressing videos efficiently. \par
Over the past several decades, a series of traditional video coding standards have been developed by the coding experts from ITU-T Video Coding Experts Group (VCEG) and ISO/IEC Motion Picture Experts Group (MPEG), such as H.264/AVC~\cite{wiegand2003overview}, H.265/HEVC~\cite{sullivan2012overview}, H.266/VVC~\cite{bross2021overview}. Thanks to a variety of advanced coding technologies, the codecs based on these coding standards have significantly improved video compression performance. Among these techniques, temporal prediction plays a vital role in reducing temporal redundancy. \par

Various prediction coding modes exist in traditional video codecs. The decision of coding modes commonly adapts to prediction quality and reference quality. In terms of prediction quality, for example, for the regions with translational motion, such as moving vehicles and pedestrians, merge mode and advanced motion vector prediction (AMVP) mode~\cite{chien2021motion} tend to obtain higher prediction quality. For the regions with non-translational motion, such as zooming and rotation, affine mode~\cite{li2017efficient,zhang2018improved} tends to be decided. In terms of reference quality, for example, if the quality of the reference frame is low, the proportion of skip mode~\cite{sullivan2012overview,bross2021overview} will decrease.\par

Although traditional video codecs are still being refined, it is more and more challenging to achieve large improvements under limited coding complexity increases. To break through the bottleneck of compression performance, in the past few years, various learned video compression schemes on top of deep neural networks have been proposed. These schemes can roughly divided into five classes: volume coding-based~\cite{Habibian_2019_ICCV,sun2020high}, temporal entropy modeling-based~\cite{liu2020conditional,DBLP:conf/nips/MentzerTMCHLA22}, 
implicit neural representation-based~\cite{chen2021nerv,chen2023hnerv,zhao2023dnerv,kwan2024hinerv}, residual coding-based~\cite{liu2020learned,Rippel_2021_ICCV,hu2020improving,lu2020content,lu2020end,lin2020m,hu2021fvc,yang2020learning,agustsson2020scale,cheng2019learning,rippel2019learned,djelouah2019neural,yang2021learning,wu2018video,liu2021deep,liu2020neural,yilmaz2021end,chen2019learning,lin2022dmvc,guo2023learning,guo2023enhanced,yang2022advancing}, and conditional coding-based~\cite{ladune2020modenet,brand2024conditional,chen2024maskcrt,sheng2022temporal,wang2023butterfly,sheng2024spatial,sheng2024vnvc,li2021deep,li2022hybrid,ho2022canf,jin2023learned,lin2023multiple,li2023neural}. 
Among them, conditional coding-based schemes have achieved the highest compression performance and some recently proposed schemes~\cite{sheng2024spatial,li2023neural,li2024neural} even have outperformed the reference software of H.266/VVC. \par
\IEEEpubidadjcol

However, different from traditional video codecs, existing conditional coding-based learned video codecs ignore the adaptation for prediction quality. They commonly use multi-channel motion compensation to learn temporal contexts as predictions. Since the complexities of video content and motion patterns are unevenly distributed in space, spatial-wise prediction quality difference exists in temporal contexts. In addition, since each channel of a temporal context is compensated by different motion vectors~\cite{li2023neural}, channel-wise prediction quality difference also exists in temporal contexts. Without explicit discrimination, the temporal contexts are directly stacked into the contextual encoder-decoder as conditions to reduce temporal redundancy. However, due to the prediction quality difference, it is difficult for the codec to decide which spatial or channel location of contexts to use as predictions.  Therefore, in this paper, we propose a simple yet effective prediction quality adaptation (PQA) module to adaptively explore temporal contexts. In this module, by comparing a temporal context and an intermediate feature of the contextual encoder or decoder, a confidence value is calculated for each spatial and channel location of the temporal context that indicates the correctness of the temporal context. As shown in Section~\ref{ablation:PQA}, with the confidence maps, the prediction with low quality will be suppressed and the prediction with high quality will be enhanced. The codec can adaptively decide which spatial or channel location of the temporal context to use as predictions. \par

In addition to prediction quality adaptation, reference quality adaptation is another essential point.  Different from traditional video codecs whose reconstruction distortion only comes from quantization, the reconstruction distortion of learned video codecs also comes from lossy non-linear transform networks. The non-linear transform networks generate an implicit quantization for input frames. The degree of the implicit quantization depends on the $\lambda$ in the rate-distortion loss function and the quality of input reference that is used to predict temporal contexts. The transform networks need to adapt to reference qualities to achieve the target reconstruction quality controlled by $\lambda$. Without considering reference quality adaptation, it is easier for learned video codes suffering from reconstruction error propagation. Existing learned video compression schemes commonly increase the length of training sequences to achieve this goal. They make transform networks ``see" a wide range of reference qualities during training, thus covering more reference qualities during testing. For example, DCVC-FM~\cite{li2024neural} uses a long-sequence (32 frames) cascaded training strategy to provide more reference qualities. However, this method consumes a large GPU memory cost. Different from previous schemes, in this work, we focus on increasing the adaptation ability of transform networks even if we use less training frames. Therefore, we propose a reference quality adaptation (RQA) module. This module learns spatially variant filters from reference frames. These dynamic filters are then applied to the intermediate features of transform networks to conduct spatially variant filtering. Given these filters, the transform networks can adapt to different reference qualities effectively. To make the RQA module ``see" different reference qualities during training, we also propose a repeat-long training strategy that combines repeating compressing and long-sequence cascaded training.  As shown in Section~\ref{ablation:RQA}, even if we only use 19 training frames, which is much smaller than that of DCVC-FM, we can still effectively mitigate the error propagation. \par
Our contributions are summarized as follows:
\begin{itemize}
    \item We propose a confidence-based prediction quality adaptation (PQA) module to adapt to the spatial and channel-wise prediction quality difference of temporal contexts.

    \item We propose a reference quality adaptation (RQA) module and an associated repeat-long training strategy to adapt to different qualities of reference, thus reducing the reconstruction error propagation.


\item Experimental results verify that our proposed modules can effectively help our codec achieve a higher compression performance.

\end{itemize}
The remainder of this paper is organized as follows.  Section~\ref{sec:related_work} gives a review of related work. Section~\ref{sec:overview} gives an overview of the learned video compression framework based on our proposed method. Section~\ref{sec:methodology} descibes our proposed methods in detail. Section~\ref{sec:experiments} presents the experimental results and ablation studies. Section~\ref{sec:conclusion} gives a conclusion of this paper.

\section{Related Work}\label{sec:related_work}
Existing learned video compression schemes can be roughly divided into the following five classes.\par
Habibian et al.~\cite{Habibian_2019_ICCV} proposed the pioneer of volume coding-based schemes. They used 3D convolution to capture the temporal correlation between multiple frames. These frames consist of a 3D volume and are compressed into a compact 3D latent code with a 3D auto-encoder. To reduce high computing costs caused by traditional 3D convolution, Sun et al.~\cite{sun2020high} proposed a frame-based 3D convolution for efficient multi-frame fusion. Currently, the compression ratio of this kind of scheme is only comparable with the industrial software of H.265/HEVC---x265 with very fast preset.\par

Liu et al.~\cite{liu2020conditional} proposed the first work of temporal entropy modeling-based schemes. They first transformed each frame into the latent space independently. When estimating the probability distribution parameters of the current latent code, they built a temporal entropy model and used the latent codes of previous frames as priors to reduce temporal redundancy. Based on this work, Mentzer et al.~\cite{DBLP:conf/nips/MentzerTMCHLA22} proposed to split the latent codes into a sequence of tokens. Then, they proposed a Transformer-based temporal entropy model to use the transmitted tokens to predict the current token. Without motion-compensated prediction, the coding complexity of this kind of scheme is smaller. Currently, the compression performance of this kind of scheme is higher than x265 with very slow preset.\par

Chen et al.~\cite{chen2021nerv} proposed the first work of implicit neural representation-based learned video compression schemes---NeRV. For video encoding, they fed the frame indexes into a neural network and fitted neural networks to regress video frames. Then, pruning, quantization, and entropy coding are performed on the parameters of neural networks. The decoding process is a simple neural network feedforward operation. To improve the video representation ability, Kwan et al.~\cite{kwan2024hinerv} further proposed to split each video frame into patches and feed the indexes of patches in different frames into neural networks. Chen et al.~\cite{chen2023hnerv} proposed to replace indexes with content-adaptive embeddings learned from video frames. The embeddings and parameters are both transmitted to the decoder. Feeding the transmitted embeddings into networks can obtain reconstructed frames. Based on this work, Zhao et al.~\cite{zhao2023dnerv} further proposed to learn a content embedding from the current frame and learn another difference embedding from the difference between the current frame and adjacent frames. The learned two embeddings are used to reconstruct videos. Because of the need to fit all video frames, the encoding time of this kind of scheme is long. However, their decoding time is short since the transmitted decoder networks are lightweight. Currently, the best compression performance~\cite{kim2024c3} of this kind of scheme is higher than x265 with very slow preset. \par

Lu et al.~\cite{lu2020end} proposed the pioneer of residual coding-based schemes---DVC. They followed the traditional hybrid video coding framework and used neural networks to implement most coding modules, such as motion estimation, motion compression, motion compensation, residue compression, and entropy models. Based on DVC, a series of work  emerged~\cite{lu2020end,yilmaz2021end,hu2022coarse,liu2020learned,Rippel_2021_ICCV,hu2020improving,lu2020content,lin2020m,hu2021fvc,yang2020learning,agustsson2020scale,cheng2019learning,rippel2019learned,djelouah2019neural,yang2021learning,wu2018video,liu2021deep,liu2020neural,liu2022end,yilmaz2021end,chen2019learning,lin2022dmvc,guo2023learning,guo2023enhanced,yang2022advancing} . Most of them focused on improving the temporal prediction efficiency. Lin et al.~\cite{lin2020m} utilized multiple reference frames and associated multiple motion vectors to generate more accurate temporal prediction of the current frame and motion vector prediction. Agustsson et al.~\cite{agustsson2020scale} replaced the commonly-used space flow with a scale-space flow which added another scale dimension to handle disocclusions and fast motion. Hu et al.~\cite{hu2020improving} proposed a resolution-adaptive flow coding method to effectively compress motion flows with multi-resolution flow representations. Hu et al.~\cite{hu2021fvc} further proposed to perform coding operations in feature space. They used the learnable offsets of deformable convolution to represent motion and deformable convolution to perform motion compensation. Currently, this kind of scheme has outperformed the reference software of H.265/HEVC---HM.\par
\begin{figure*}[t]
  \centering
   \includegraphics[width=0.75\linewidth]{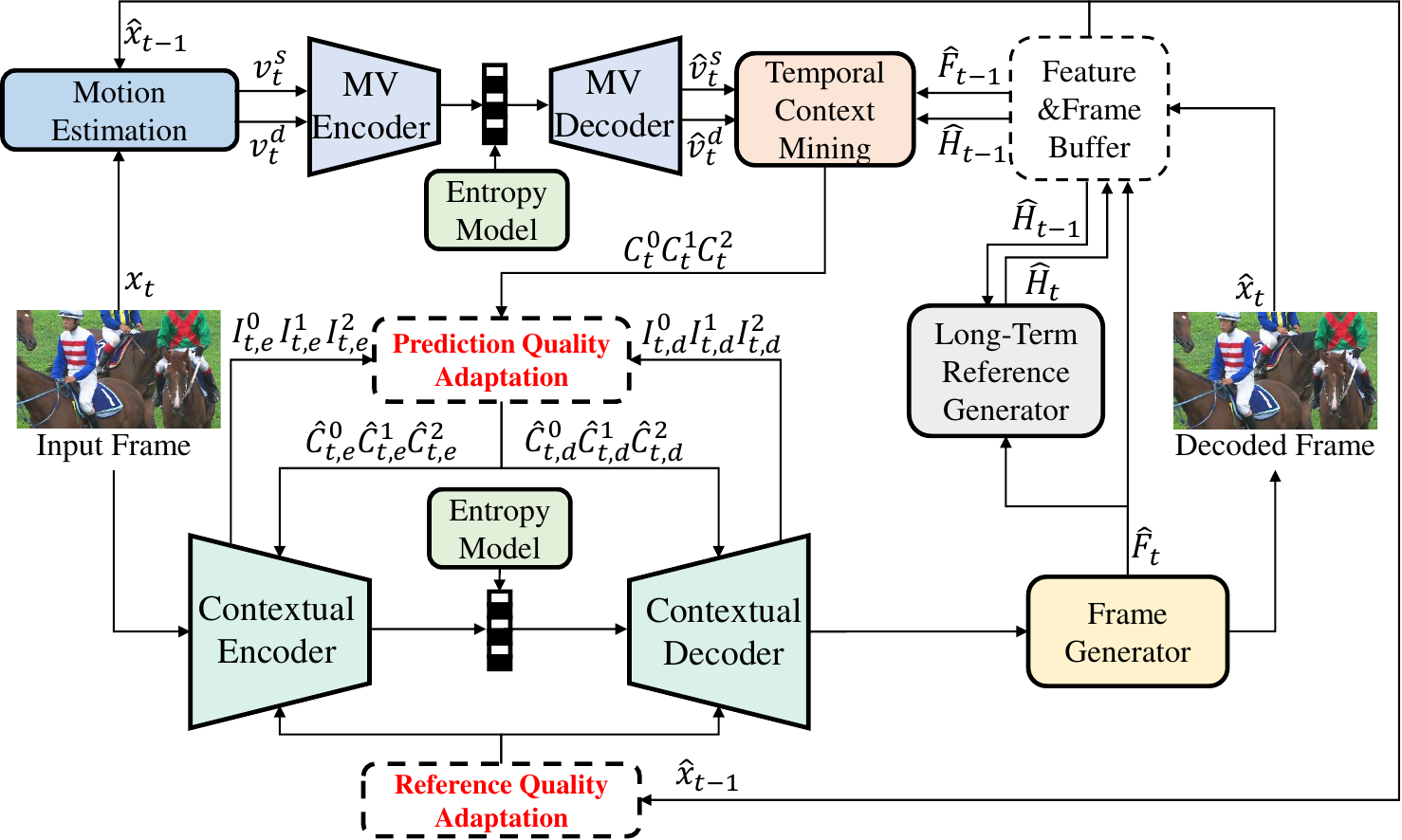}
      \caption{Overview of the learned video codec based on our proposed prediction quality adaptation module and reference quality adaptation module.}
   \label{fig:framework}
\end{figure*}
\begin{figure*}[t]
  \centering
   \includegraphics[width=0.85\linewidth]{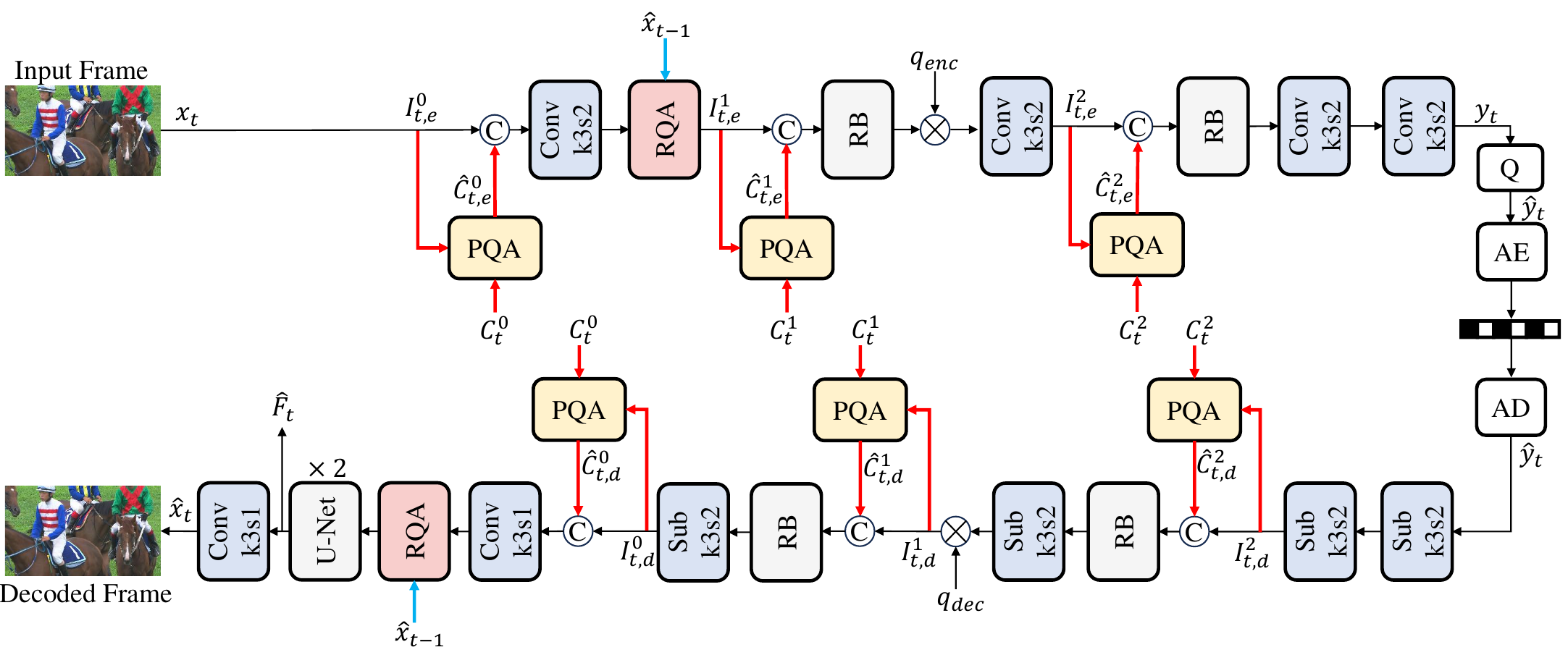}
      \caption{Architecture of the contextual encoder-decoder and frame generator. The proposed prediction quality adaptation (PQA) module is used to provide explicit spatial and channel-wise discrimination for the prediction quality of temporal contexts. The proposed reference quality adaptation (RQA) module is used to dynamically adjust the contextual encoder-decoder to adapt to reference quality to reduce reconstruction error propagation.}
   \label{fig:contextual_encoder_decoder}
\end{figure*}
Li et al.~\cite{li2021deep} proposed the representative conditional coding-based scheme---DCVC. They performed motion compensation to the feature learned from a reference frame to generate a single-scale temporal context. Rather than calculating the residue, they regarded the temporal context as a condition and concatenated it with the input frame. The contextual encoder and decoder can learn how to take advantage of the temporal context to reduce temporal redundancy sufficiently. Based on DCVC, Sheng et al.~\cite{sheng2022temporal} proposed DCVC-TCM, which uses reference features instead of reference frames to propagate the temporal information. From the reference features, they designed a temporal context mining (TCM) module to predict multi-scale temporal contexts and proposed a temporal context re-filling (TCR) method to make full use of the predicted multi-scale temporal contexts. Inheriting the TCM and TCR in DCVC-TCM, most extended schemes~\cite{sheng2022temporal,li2021deep,li2022hybrid,li2023neural, sheng2024spatial,wang2023butterfly} focus on generating more accurate temporal contexts. Sheng et al.~\cite{sheng2024spatial} proposed a spatial decomposition-based motion model and a long-term temporal fusion module to handle motion inconsistency and motion occlusion. Li et al.~\cite{li2023neural,li2024neural} design a hierarchical quality structure and a feature refreshing method to periodically generate high-quality temporal contexts. Currently, this kind of scheme has outperformed the reference software of H.266/VVC---VTM.\par

\section{Overview}\label{sec:overview}
To adapt to temporal contexts with different qualities caused by prediction quality difference and reference quality difference, we propose a temporal context quality adaptation method based on our baseline---DCVC-SDD~\cite{sheng2024spatial}. We first give an overview to introduce the framework based on our proposed methods. 

\subsubsection{Motion Estimation}To estimate the motion between video frames, we use the structure and detail decomposition-based motion estimation method proposed in our baseline~\cite{sheng2024spatial}. The input frame $x_t$ and reference frame $\hat{x}_{t-1}$ are first decomposed into structure and detail components. Then, the motion vectors ($v_{t}^{s}$, $v_{t}^{d}$) between their structure and detail components are estimated respectively using SpyNet~\cite{ranjan2017optical}. 

\subsubsection{Motion Coding}After obtaining the motion vectors of structure and detail components, a motion encoder compresses $v_{t}^{s}$ and $v_{t}^{d}$ jointly into a compact latent representation $\hat{m}_t$. After receiving the transmitted latent representation $\hat{m}_t$, the motion decoder inversely transforms it back to the reconstructed motion vectors $\hat{v}_{t}^{s}$ and $\hat{v}_{t}^{d}$. Similar to~\cite{li2023neural, sheng2024spatial}, a learnable motion quantization step and an inverse motion quantization step is embedded into the motion encoder and decoder respectively for supporting rate adjustment with a single model.

\subsubsection{Temporal Context Mining}Following our baseline~\cite{sheng2024spatial}, we use a structure and detail decomposition-based temporal context mining module~\cite{sheng2022temporal, sheng2024spatial,li2022hybrid,li2023neural} to perform feature-domain motion compensation to the reference feature $\hat{F}_{t-1}$ to generate short-term multi-scale temporal contexts $\bar{C}_t^{0}, \bar{C}_t^{1}, \bar{C}_t^{2}$. To handle motion occlusion, we fuse a long-term reference feature $\hat{H}_{t-1}$ with $\bar{C}_t^{0}, \bar{C}_t^{1}, \bar{C}_t^{2}$ to generate long short-term fused temporal contexts $C_t^{0}, C_t^{1}, C_t^{2}$.

\subsubsection{PQA and RQA-Enhanced Contextual Encoder-Decoder}
The contextual encoder and decoder are based on an auto-encoder structure as shown in Fig.~\ref{fig:contextual_encoder_decoder}. The input frame $x_t$ is transformed into a compact latent representation $y_t$. After quantization, the latent representation $\hat{y}_t$ is signaled into the bitstream by the arithmetic encoder (AE). After receiving the bitstream at the decoder side, the arithmetic decoder (AD) performs entropy decoding to the bitstream to obtain $\hat{y}_t$. Then, the contextual decoder and frame generator inversely transform $\hat{y}_t$ back to the reconstructed frame $\hat{x}_t$. We regard the input feature $\hat{F}_t$ of the last convolutional layer of the frame generator as the reference feature of the next frame $x_{t+1}$. Similar to the motion encoder and decoder, a learnable contextual quantization step and an inverse contextual quantization step are integrated into the contextual encoder and decoder, respectively, to achieve a variable rate. Following the paradigm of temporal context mining-based scheme~\cite{li2023neural, sheng2022temporal, sheng2024spatial}, the learned multi-scale temporal contexts $C_t^{0}, C_t^{1}, C_t^{2}$ are fed into the contextual encoder and decoder as conditions~\cite{li2021deep}. Considering that spatial and channel-wise prediction quality differences exist in temporal contexts, we propose a prediction quality adaptation module (PQA) to provide explicit discrimination for the quality of temporal contexts $C_t^{0}, C_t^{1}, C_t^{2}$. After discrimination, the refined temporal contexts $\hat{C}_t^{0}, \hat{C}_t^{1}, \hat{C}_t^{2}$ serve as new conditions. In addition to prediction quality, reference quality also needs to be adapted by the contextual encoder and decoder to reduce reconstruction error propagation. Therefore, we propose a reference quality adaptation module (RQA) to dynamically adjust the contextual encoder and decoder to adapt to different reference qualities. More details about PQA and RQA will be described in Section~\ref{sec:PQA} and Section~\ref{sec:RQA}.

\subsubsection{Frame Generator}
After obtaining the feature decoded by the contextual decoder, we use a frame generator which is comprised of two U-Nets~\cite{sheng2024spatial,li2022hybrid,li2023neural} reconstructs the feature back to the pixel-domain reconstructed frame $\hat{x}_{t}$. Before obtaining $\hat{x}_{t}$, we regard the input feature $\hat{F}_t$ of the last convolutional layer as the reference feature of the next frame $x_{t+1}$.

\subsubsection{Long-Term Reference Generator}
To handle motion occlusion, we follow our baseline~\cite{sheng2024spatial} and introduce a ConvLSTM-based long-term reference generator. After obtaining the reference feature $\hat{F}_t$ of the next frame $x_{t+1}$, we feed $\hat{F}_t$ into the long-term reference generator to generate a long-term reference feature $\hat{H}_t$, which will be used in the temporal context mining module to provide long-term temporal contexts.

\subsubsection{Entropy Model}
We use hyperprior entropy model~\cite{DBLP:conf/iclr/BalleMSHJ18}, quadtree partition-based spatial entropy model~\cite{li2023neural}, and conditional temporal entropy model~\cite{sheng2022temporal,sheng2024spatial,li2022hybrid,li2023neural} to jointly estimate the probability distribution of motion vector latent representation $\hat{m}_t$ and contextual latent representation $\hat{y}_t$, which are both assumed to follow the Laplace distribution.

\section{Methodology}\label{sec:methodology}
\subsection{Prediction Quality Adaptation}\label{sec:PQA}
\begin{figure}[t]
  \centering
   \includegraphics[width=0.9\linewidth]{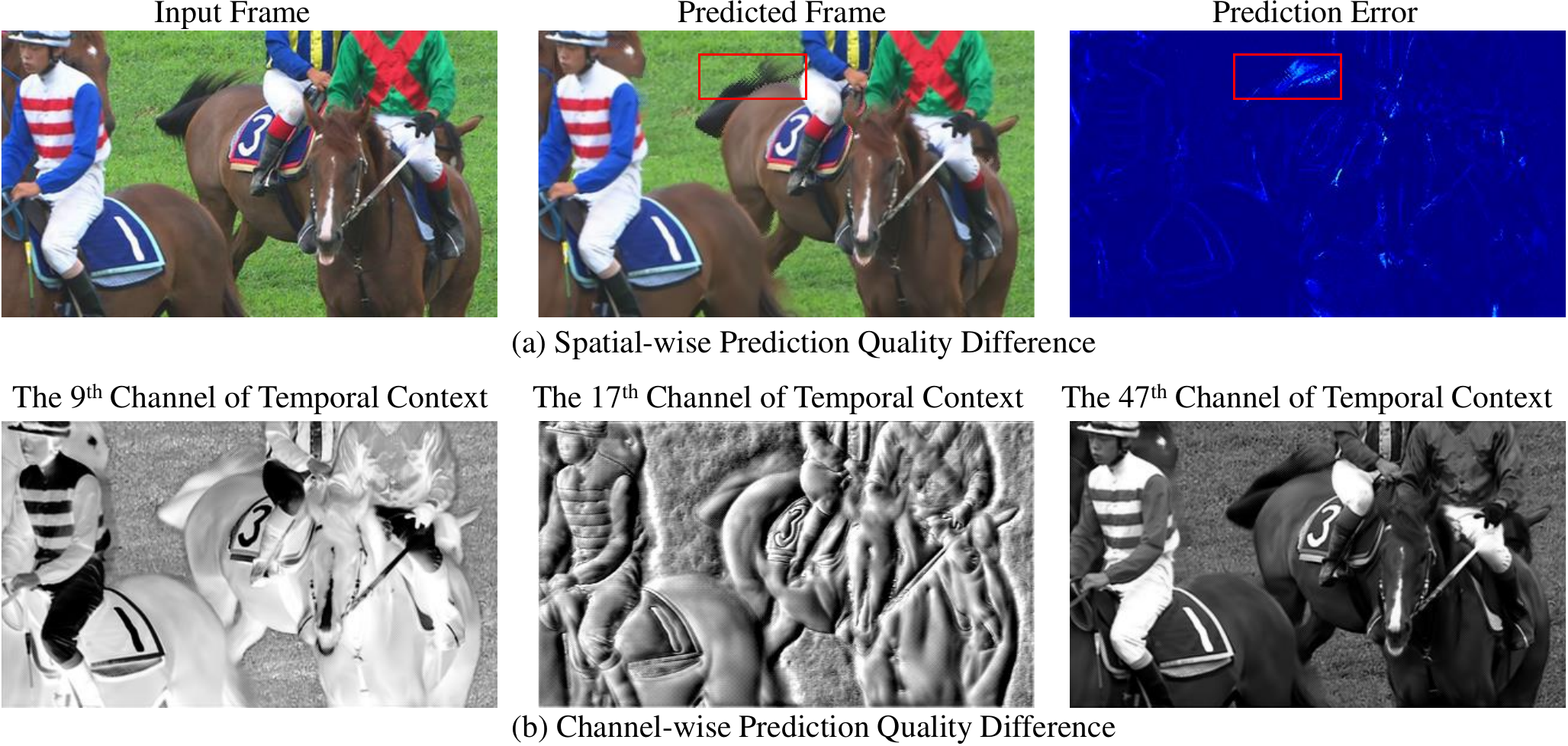}
      \caption{Visualization of the spatial and channel-wise prediction quality difference of temporal contexts.}
\label{fig:PQA_example}
\end{figure}

Temporal contexts are generated by feature-domain temporal prediction. Therefore, the quality of temporal prediction affects the temporal context quality. Since the complexities of video content and motion patterns are unevenly distributed in space, spatial-wise prediction quality difference exists in temporal contexts. For example, as shown in Fig.~\ref{fig:PQA_example} (a), the prediction error of the horsetail in the red rectangle is large since it has complex textures and motion, while that of the horse body is smaller since their textures are simpler and their motion pattern is mainly simple translational motion. Using temporal context with large spatial prediction errors may increase coding costs. Therefore, it is necessary to give explicit discrimination to the spatial-wise temporal context quality difference to help the context encoder and decoder decide whether to use the temporal context in a certain region.\par
In addition to the spatial-wise prediction quality difference, channel-wise prediction quality difference also exists in temporal contexts. Commonly, temporal contexts have multiple feature channels, and each channel is compensated by different motion vectors~\cite{li2023neural}. Therefore, temporal information is unevenly distributed across context channels. Different context channels may have different prediction qualities and contain different temporal information. For example, as shown in Fig.~\ref{fig:PQA_example} (b), the $9^{th}$ and $47^{th}$ channels of the temporal context have more temporal information and their temporal prediction qualities are higher, while the $17^{th}$ channel of temporal context has less temporal information and its temporal prediction quality is lower. Therefore, it is also necessary to give explicit discrimination to the channel-wise temporal context quality difference to help the contextual encoder and decoder decide which channel of temporal information to use.\par

\begin{figure}[t]
  \centering
   \includegraphics[width=0.95\linewidth]{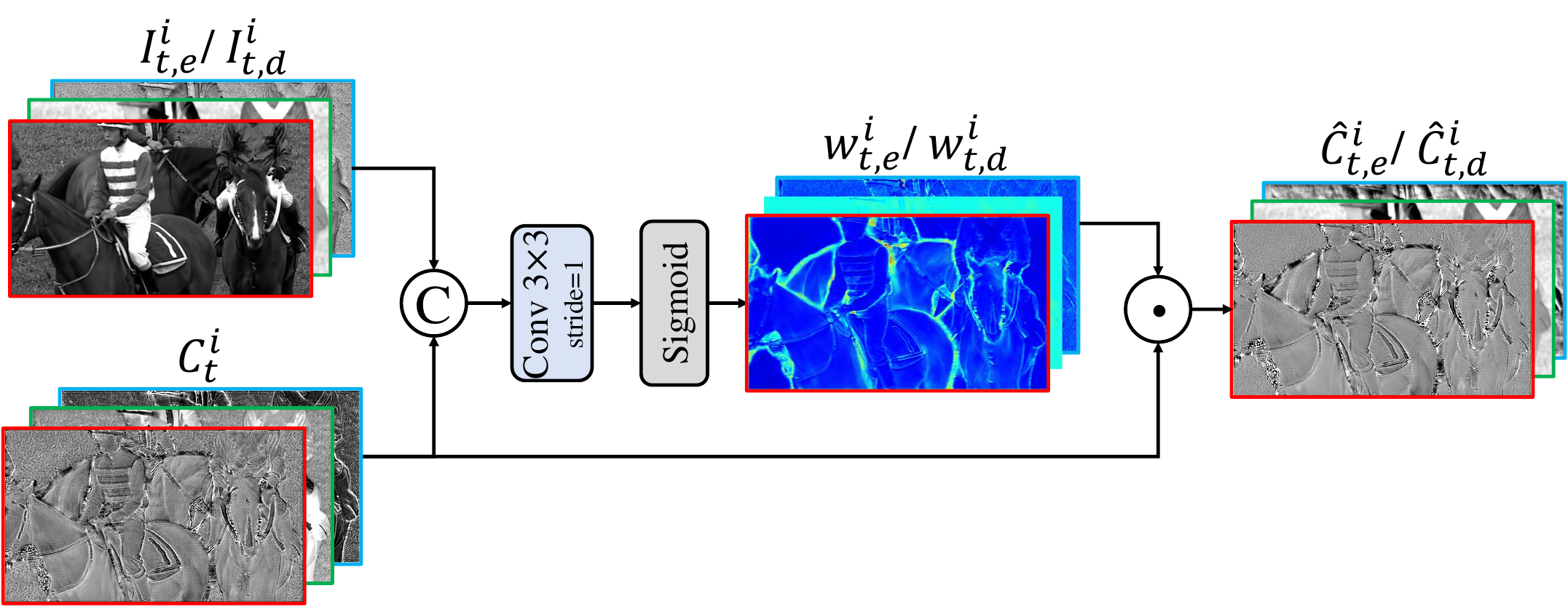}
      \caption{Architecture of our proposed prediction quality adaptation module.}
\label{fig:PQA}
\end{figure}

To give explicit discrimination to the spatial and channel-wise prediction quality difference of temporal contexts, we design a simple yet effective prediction quality adaptation module as illustrated in Fig.~\ref{fig:PQA}.
Given a certain scale temporal context $C_{t}^i$, we first concatenate it with an intermediate feature $I_{t,e}^i$ of the contextual encoder or an intermediate feature $I_{t,d}^i$ of the contextual decoder with the same scale. For the biggest temporal context $C_t^0$ in the encoder, the input frame $x_t$ serves as $I_{t,e}^0$. Then, we feed them into a convolutional layer to obtain a feature with the same spatial and channel dimension as $C_t^i$. The feature is passed into a Sigmoid function to obtain a group of confidence maps for each channel of the temporal context $C_t^i$. Considering that the temporal contexts for reducing temporal redundancy and video reconstruction may be different, given a temporal context $C_t^{i}$, we calculate different confidence maps for the contextual encoder and decoder.
\begin{equation}
\begin{aligned}
w_{t,e}^{i}&=\sigma\left(\mathcal{F}_e\left(C_t^{i} \textcircled{c} I_{t,e}^{i}\right)\right), \\
w_{t,d}^{i}&=\sigma\left(\mathcal{F}_d\left(C_t^{i} \textcircled{c} I_{t,d}^{i}\right)\right), \\
i &= 0,1,2,
\end{aligned}
\end{equation}
where \textcircled{c} is the concatenation operation. $w_{t,e}^{i}$ is the confidence map group for contextual encoder and $w_{d,t}^{i}$ is the confidence map group for contextual decoder. $i$ is the scale index of temporal contexts. $\mathcal{F}_e$ and $\mathcal{F}_d$ denote 2D convolution operations with $3 \times 3$ filter size. $\sigma$ refers to the Sigmoid function. Each element of $w_{t,e}^{i}$ and $w_{d,t}^{i}$ is in the range of 0 to 1. The greater the value of one element, the more important the temporal context of the corresponding position is.
 \par

After obtaining the confidence map for each channel of $C_t^i$, we perform element-wise product operation between the confidence map and the corresponding channel of temporal context.
\begin{equation}
\begin{aligned}
\hat{C}_{t,e}^{i,m} &= w_{t,e}^{i,m} \odot C_{t}^{i,m} \\
\hat{C}_{t,d}^{i,m} &= w_{t,d}^{i,m} \odot C_{t}^{i,m} \\
m &= 0, \cdots, M-1,
\end{aligned}
\end{equation}
where $\odot$ is the element-wise product operation, $m$ is the channel index of the temporal context $C_t^{i}$, and $M$ is the number of channels of $C_t^{i}$. 

\subsection{Reference Quality Adaptation}\label{sec:RQA}
\begin{figure}[t]
  \centering
   \includegraphics[width=0.7\linewidth]{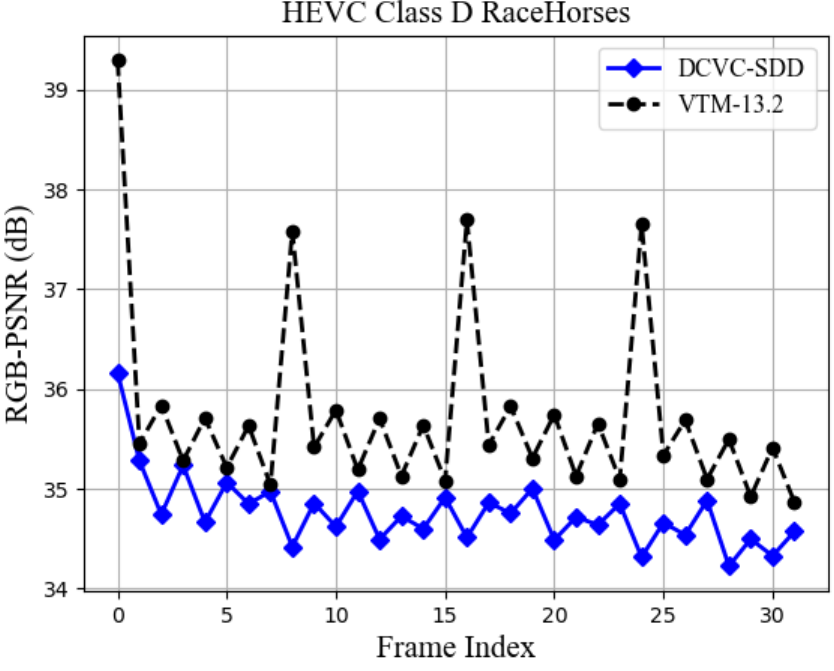}
      \caption{Illustration of the reconstruction quality difference across video frames of the reference software of H.266/VVC~\cite{bross2021overview} and our baseline---DCVC-SDD~\cite{sheng2024spatial}.}
\label{fig:RQA_example}
\end{figure}

In addition to prediction quality, reference quality also needs to be adapted to reduce reconstruction error propagation.  As shown in Fig.~\ref{fig:RQA_example}, the continuous degradation of reconstruction quality is relatively limited in traditional video codecs, such as VTM. Their transforms like DCT are lossless and their reconstruction distortion only comes from the quantization. Therefore, it is easier for them to achieve a target reconstruction quality by adjusting a quantization step. However, in learned video codecs, the reconstruction distortion not only comes from quantization but also comes from non-linear transform networks. The non-linear transform networks (\emph{$Encoder_{\theta_e}$}, \emph{$Decoder_{\theta_d}$}) generate an implicit quantization for the input frame. The degree of the implicit quantization depends on the $\lambda$ in the rate-distortion loss function and the quality of reference $\hat{x}_{t-1}$ that is used to predict temporal contexts. In other words, the transform networks will adapt to the reference quality to achieve the target reconstruction quality controlled by $\lambda$. \par

\begin{equation}
\begin{gathered}
L_t=\lambda D_t+R_t \\
D_t=MSE\left(x_t, Decoder_{\theta_d}\left(Encoder_{\theta_e}\left(x_t \mid \hat{x}_{t-1}\right) \mid \hat{x}_{t-1}\right)\right)
\end{gathered}
\end{equation}

Existing learned video compression schemes commonly increase the length of training sequences to achieve this goal. They make transform networks ``see" a wide range of reference qualities during training, thus covering more reference qualities during testing. For example, DCVC-FM~\cite{li2024neural} uses a long-sequence (32 frames) cascaded training strategy to provide more reference qualities. However, this method consumes a large GPU memory cost. Therefore, we focus on making the transform networks adapt to different reference qualities even if we use less training frames. Some schemes~\cite{li2023neural,sheng2024spatial} proposed to add a periodically varying weight before $\lambda$ in the loss function to increase the adaptation ability of transform networks. However, it is still difficult to adapt various reference qualities with a limited number of weights (1.2, 0.5, 0.9) to achieve the target quality. As shown in Fig.~\ref{fig:RQA_example}, for our baseline DCVC-SDD~\cite{sheng2024spatial}, which has used this kind of reference quality adaptive loss function, the qualities of its reconstructed frames still gradually decrease as the reference qualities decrease.\par

\begin{figure}[t]
  \centering
   \includegraphics[width=0.85\linewidth]{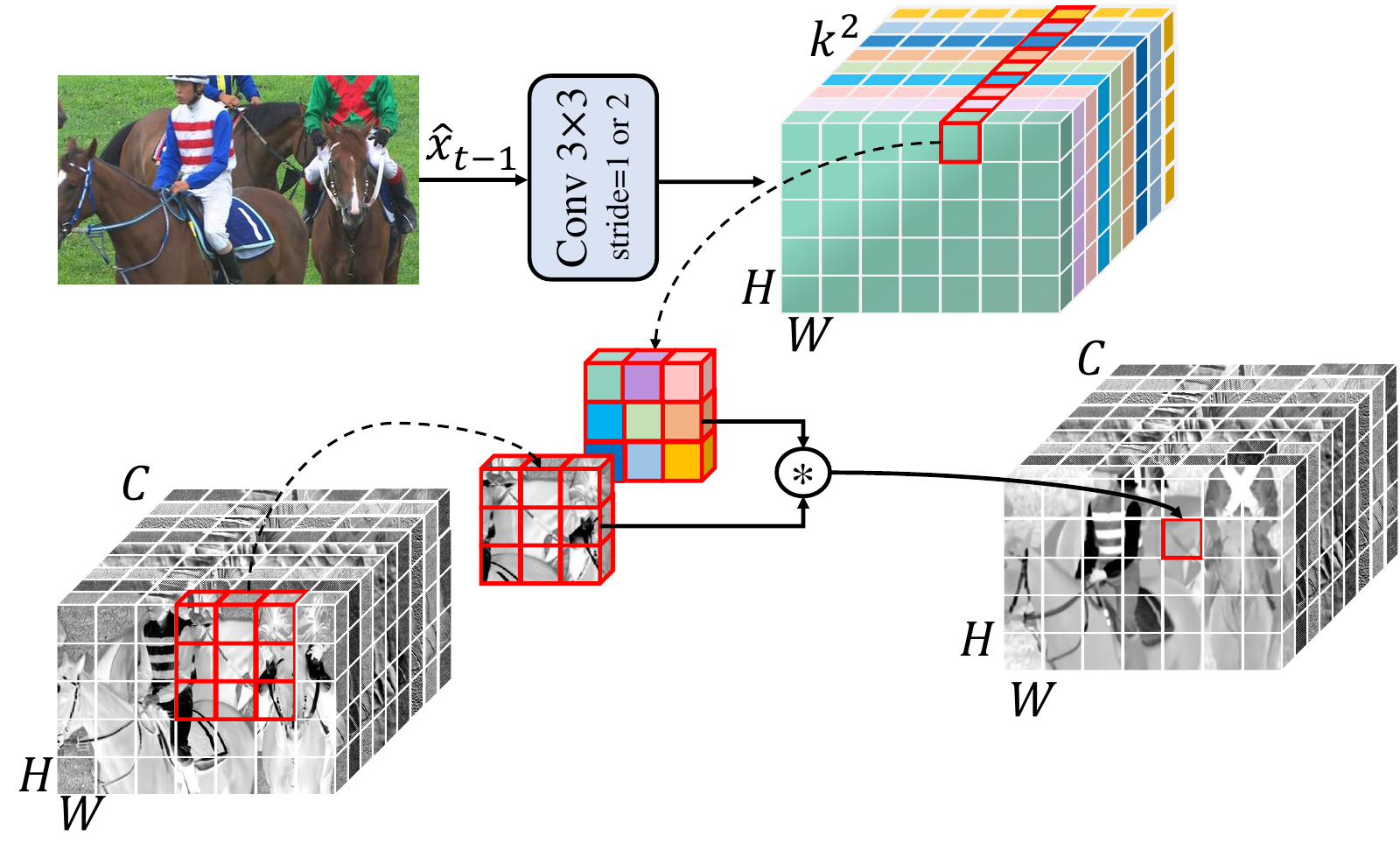}
      \caption{Architecture of our proposed reference quality adaptation module. $H$ and $W$ are height and width of the intermediate feature, and $k$ refers to the size of filters. To embed the RQA module into our baseline DCVC-SDD while keeping its original main architecture unchanged, for the RQA module in the encoder, the stride of the convolutional layer is 2, and for the RQA module in the decoder, the stride of the convolutional layer is 1.}
\label{fig:RQA}
\end{figure}
To make the transform networks adapt to reference qualities well, we propose a reference quality adaptation (RQA) module. As presented in Fig.~\ref{fig:RQA}, we first feed the reference frame $\hat{x}_{t-1}$ into a convolutional layer $\mathcal{F}_{RQA}$ to obtain spatially variant filters $W_t^{\theta}\in R^{H \times W \times k \times k}$ for each position of an intermediate feature $I_t$ of the contextual encoder or decoder. $H$ and $W$ are height and width of the intermediate feature, and $k$ refers to the size of filters: 
\begin{equation}
W_t^{\theta} = \mathcal{F}_{RQA}\left(\hat{x}_{t-1}\right).
\end{equation}
Then, each filter $W_t^{\theta}(i,j)$ is applied to a $k \times k$ window of the intermediate feature $I_t$ centered at position $(i,j)$ to conduct spatially variant filtering.
\begin{equation}
O_{t}(i, j, c)=\sum_{u=-\lceil k / 2\rfloor}^{\lfloor k / 2\rfloor} \sum_{v=-\lfloor k / 2\rfloor}^{\lfloor k / 2\rfloor} W_{t}^{\theta}(i, j, u, v) \cdot I_{t}(i+u, j+v, c).
\end{equation}
\par
To make the RQA module ``see" different reference qualities during training, we add a repeat-long training step that combines repeating compressing and long-sequence cascaded training into the commonly-used multiple-step training strategy, which will be introduced in Section~\ref{sec:training_strategy}.

\begin{table}[t]
\caption{Training strategy of our scheme for encoding RGB videos when distortion is measured by RGB PSNR.}
\centering
\begin{tabular}{c|c|c|c|c}
\toprule[1.5pt]
Frames  & Network   & Loss                          &LR        &Epoch \\ \hline
2     & Inter   & $ L_{t}^{meD}$                    &$1e-4$     & 2      \\ \hline
2     & Recon   & $ L_{t}^{recD}$                   &$1e-4$     & 1      \\ \hline
3     & Recon   & $ L_{t}^{recD}$                   &$1e-4$     & 1      \\ \hline
6     & Recon   & $ L_{t}^{recD}$                   &$1e-4$     & 1      \\ \hline
6     & Inter   & $ L_{t}^{meD}$                    &$1e-4$     & 2      \\ \hline
6     & Inter   & $ L_{t}^{meRD}$                   &$1e-4$     & 6      \\ \hline
6     & Recon   & $ L_{t}^{recD}$                   &$1e-4$     & 2      \\ \hline
6     & Recon   & $ L_{t}^{recRD}$                  &$1e-4$     & 6      \\ \hline
6     & All     & $ L_{t}^{all}$                    &$1e-4$     & 4      \\ \hline
6     & All     & $ L_{t}^{all}$                    &$5e-5$     & 3      \\ \hline
6     & All     & $ L_{t}^{all}$                    &$1e-5$     & 3      \\ \hline
6     & All     & $ L_{t}^{all}$                    &$5e-6$     & 4      \\ \hline
6     & All     & $ L_{T}^{all}$                    &$5e-5$     & 2      \\ \hline
6     & All     & $ L_{T}^{all}$                    &$5e-6$     & 2      \\ \hline
6     & All     & $ L_{T}^{all}$                    &$1e-6$     & 1      \\
\hline
19     & All    & $ L_{T}^{all-r-l}$        &$1e-6$     & 1      \\

\bottomrule[1.5pt]
\end{tabular}
\label{training_stategy_rgb}
\end{table}

\subsection{Training Strategy}\label{sec:training_strategy}
We propose a step-by-step training strategy to train our learned video compression scheme. The training details for encoding RGB videos when the reconstruction distortion is measured by RGB PSNR are listed in Table.~\ref{training_stategy_rgb}. 
According to the training loss function, the training strategy can be classified into 7 classes: $L_{t}^{meD}$, $L_{t}^{meRD}$, $L_{t}^{recD}$, $L_{t}^{recRD}$, $L_{t}^{all}$, $L_{T}^{all}$, and $ L_{T}^{all-r-l}$.
\begin{itemize}
\item \textbf{$L_{t}^{meD}$} is comprised of $D_{t}^{me}$.
We denote $D_{t}^{me}$ to the distortion between $x_t$ and its predicted frame $\tilde{x}_t$. MSE is used to measure the distortion. The predicted frame $\tilde{x}_t$ is generated by warping the reference frame $\hat{x}_{t-1}$ using the decoded motion vectors $\hat{v}_{t}$. Using $L_{t}^{meD}$ as the loss function is to make the motion decoder generate accurate motion vectors:
\begin{equation}
    L_{t}^{meD}= w_t \cdot \lambda \cdot D_{t}^{me}.
\label{loss1}
\end{equation}
\item \textbf{$L_{t}^{meRD}$} is comprised of $D_{t}^{me}$ and $R_{t}^{me}$. We denote $R_{t}^{me}$ to the joint bit rate used for encoding the quantized motion latent representation $\hat{m}_t$ and its associated hyperprior. Using $L_{t}^{meRD}$ as the loss function is to achieve a trade-off between the accuracy and the consumed bit rate of motion vectors:
\begin{equation}
    L_{t}^{meRD}= w_t \cdot \lambda \cdot D_{t}^{me} + R_{t}^{me}.
\label{loss2}
\end{equation}

\item \textbf{$L_{t}^{recD}$} is comprised of $D_{t}^{rec}$. We denote $D_{t}^{rec}$ to the distortion between $x_t$ and its reconstructed frame $\hat{x}_t$. MSE is used to measure the distortion. Using $L_{t}^{recD}$ as the loss function is to make the contextual decoder and frame generator generate a high-quality reconstructed frame:
\begin{equation}
    L_{t}^{recD}= w_t \cdot \lambda \cdot D_{t}^{rec}.
\label{loss3}
\end{equation}

\item \textbf{$L_{t}^{recRD}$} is comprised of $D_{t}^{rec}$ and $R_{t}^{rec}$. We denote $R_{t}^{rec}$ to the joint bit rate used for encoding the quantized contextual latent representation $\hat{y}_t$ and its associated hyperprior of $x_t$. Using $L_{t}^{recRD}$ as the loss function is to achieve a trade-off between the quality of reconstructed frame $\hat{x}_t$ and the consumed bit rate of contextual latent representation $\hat{y}_t$:
\begin{equation}
    L_{t}^{recRD}= w_t \cdot \lambda \cdot D_{t}^{rec} + R_{t}^{rec}.
\label{loss4}
\end{equation}

\item \textbf{$L_{t}^{all}$} is comprised of $D_{t}^{rec}$ and $R_{t}^{all}$. We denote $R_{t}^{all}$ to all the bit rates used for encoding the quantized contextual latent representation $\hat{y}_t$, the quantized motion latent representation $\hat{m}_t$, and their hyperprior. Using $L_{t}^{all}$ as the loss function is to achieve a trade-off between the quality of the reconstructed frame and all the consumed bit rates of the coded frame:
\begin{equation}
\begin{aligned}
    L_{t}^{all}&= w_t \cdot \lambda \cdot D_{t}^{rec} + R_{t}^{all} \\
&= w_t \cdot \lambda \cdot D_{t}^{rec} + R_{t}^{me} + R_{t}^{rec}.
\end{aligned}
\label{loss5}
\end{equation}

\item \textbf{$L_{T}^{all}$} is the average $L_{t}^{all}$ loss of $T$ consecutive frames. Calculating the average loss of multiple frames is to achieve a cascaded fine-tuning for reducing the error propagation~\cite{sheng2022temporal,sheng2024spatial,li2022hybrid,li2023neural}:
\begin{equation}
\begin{aligned}
L_{T}^{all}&=\frac{1}{T} \sum_t L_t^{all}\\
&=\frac{1}{T} \sum_t\left\{w_t \cdot \lambda \cdot D_{t}^{rec} + R_{t}^{all}\right\}\\
&=\frac{1}{T} \sum_t\left\{w_t \cdot \lambda \cdot D_{t}^{rec} + R_{t}^{me} + R_{t}^{rec}\right\}.
\end{aligned}
\label{loss6}
\end{equation}
\begin{figure}[t]
  \centering
   \includegraphics[width=0.9\linewidth]{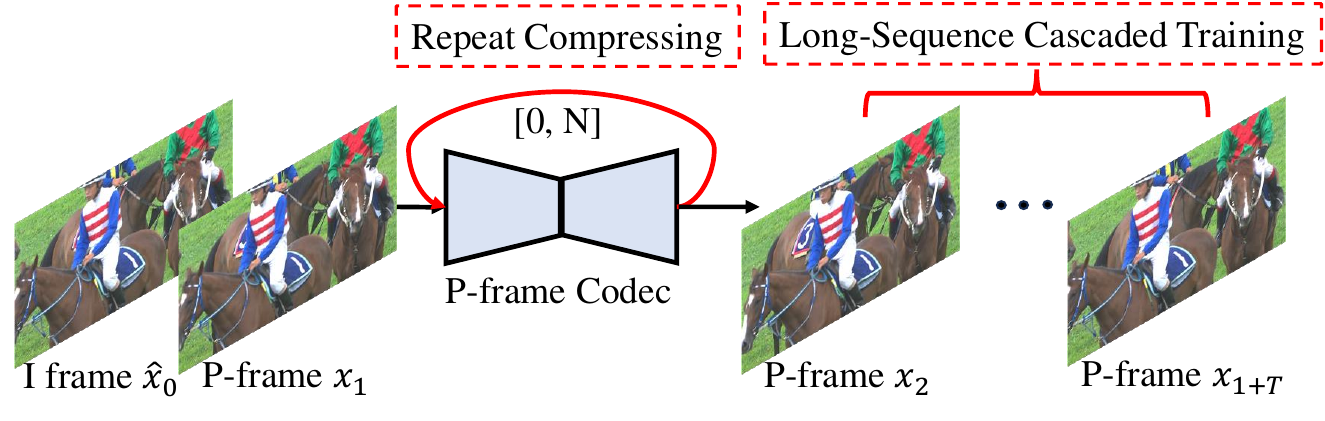}
      \caption{Illustration of our proposed repeat-long training step that combines the repeating compressing and long-sequence cascaded training.}
\label{fig:repeat_long_loss}
\end{figure}

\item \textbf{$L_{T}^{all-r-l}$} is the RQA-associated repeat-long training loss proposed in our paper to make RQA module ``see" more reference qualities during training, as described in Section~\ref{sec:RQA}.
To generate reference frames with various qualities, we first repeat compressing the first P-frame for certain times (randomly selected from 0 to $N$). Then, we regard its reconstructed frame as the reference of the second P-frame and calculate the cascaded loss $L_{T}^{all}$ of the remaining $T$ frames. A similar repeating compressing strategy is proposed by the coding experts of~\cite{shi2022alphavc} and has been accepted into the developing AVS end-to-end video coding standard reference model AVS-EEM-v2.0\footnote{AVS video coding group is exploring a lightweight end-to-end learned video coding standard. AVS-EEM is the corresponding reference model. Its training and testing codes can be accessed at https://gitlab.com/xhsheng/avs-eem once authorized. }. However, we find that directly applying the repeating compressing strategy only brings a little compression performance improvement in our model. This is because this strategy can only generate reference with various qualities for the first P-frame, and cannot affect the reference qualities of subsequent P-frames. Therefore, we combine this strategy with long-sequence training. We increase the number of frames $T$ ($T=17$) used to calculate the cascaded loss $L_{T}^{all}$ to generate various reference qualities for subsequent P-frames. With the repeating compressing and long-sequence training strategies, the RQA module can adapt to a wider range of reference qualities.
\end{itemize}

\begin{table}[t]
\caption{Fine-tuning Strategy of Our Scheme for encoding RGB videos when distortion is measured by MS-SSIM.}
\centering
\begin{tabular}{c|c|c|c|c}
\toprule[1.5pt]
Frames  & Network   & Loss                          &LR        &Epoch \\ \hline
6     & All     & $ L_{T}^{all}$                    &$5e-5$     & 4      \\
\hline
6     & All    & $ L_{T}^{all}$                     &$5e-6$     & 2      \\
\hline
6     & All    & $ L_{T}^{all}$                     &$1e-6$     & 1      \\
\hline
19     & All    & $ L_{T}^{all-r-l}$        &$1e-6$     & 1      \\
\bottomrule[1.5pt]
\end{tabular}
\label{training_stategy_ssim}
\end{table}
\begin{table}[t]
\caption{Fine-tuning strategy of our scheme for encoding YUV420 videos when distortion is measured by YUV PSNR..}
\centering
\begin{tabular}{c|c|c|c|c}
\toprule[1.5pt]
Frames  & Network   & Loss                          &LR        &Epoch \\ \hline
6     & All     & $ L_{t}^{all}$                    &$1e-4$     & 4      \\ \hline
6     & All     & $ L_{t}^{all}$                    &$5e-5$     & 4      \\
\hline
6     & All     & $ L_{t}^{all}$                    &$5e-6$     & 1      \\
\hline
6     & All    & $ L_{T}^{all}$                     &$5e-5$     & 2      \\
\hline
6     & All    & $ L_{T}^{all}$                     &$5e-6$     & 2      \\
\hline
6     & All    & $ L_{T}^{all}$                     &$1e-6$     & 1      \\
\hline
19     & All    & $ L_{T}^{all-r-l}$                     &$1e-6$     & 1      \\
\bottomrule[1.5pt]
\end{tabular}
\label{training_stategy_yuv}
\end{table}
\par
\begin{figure*}[t]
  \centering
  \begin{minipage}[c]{\linewidth}
  \centering
  \includegraphics[width=0.9\linewidth]{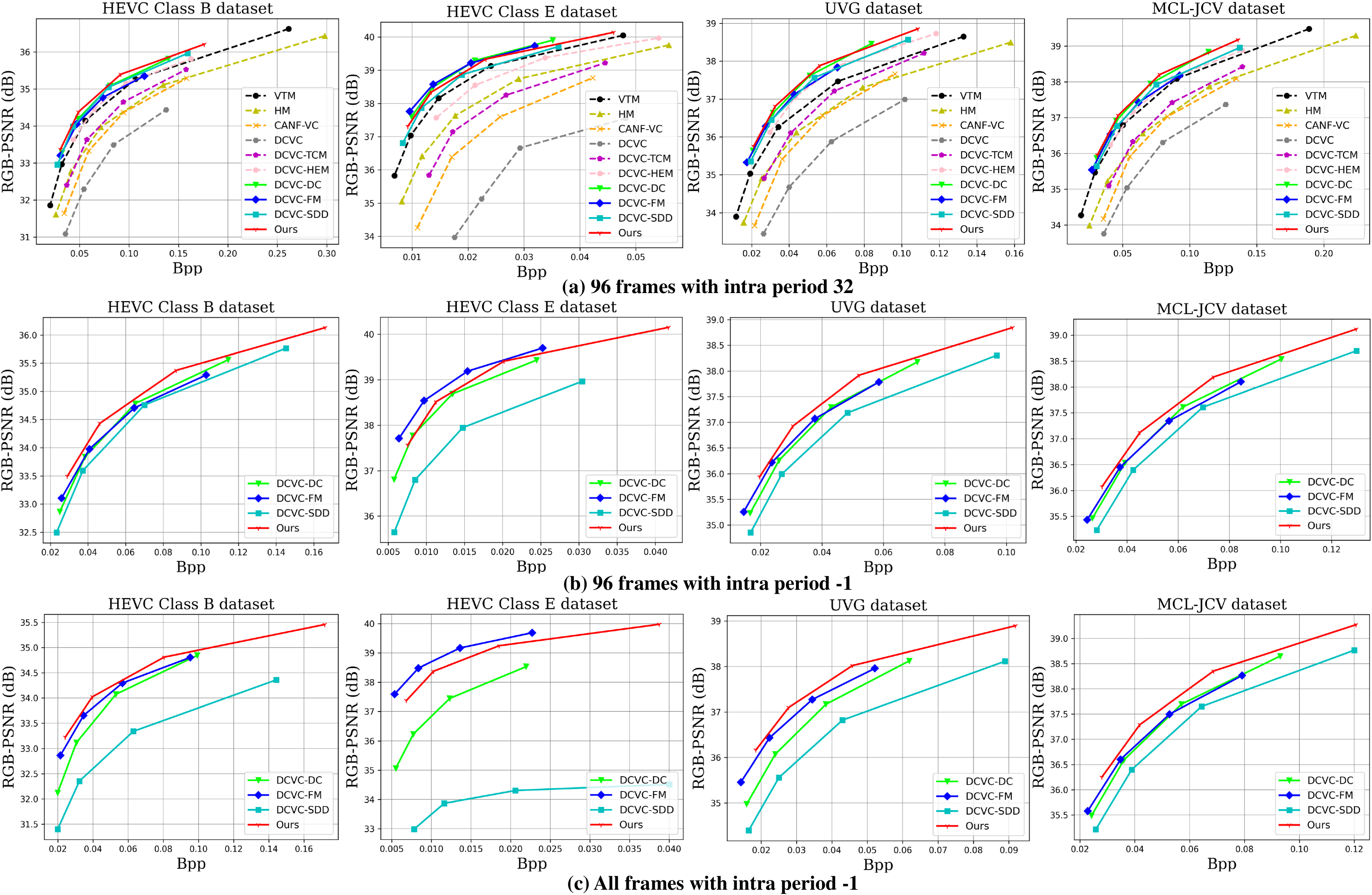}
 \end{minipage}%
    \caption{Rate-distortion curves when the quality is measured by RGB PSNR. }
  \label{fig:psnr_results}
\end{figure*}
\begin{table*}[!htb]
\caption{BD-rate(\%) comparison in RGB colorspace measured with PSNR. The anchor is VTM. \textbf{96 frames with intra period is 32.}}
  \centering
\scalebox{1}{
\begin{tabular}{l|c|c|c|c|c|c|c}
\toprule[1.5pt]
               & HEVC Class B  & HEVC Class C  &HEVC Class D &HEVC Class E &HEVC Class RGB &UVG            &MCL-JCV  \\ \hline
VTM            &0.0            &0.0            &0.0          &0.0          &0.0            &0.0             &0.0      \\ \hline
HM             &39.0           &37.6           &34.7         &48.6         &44.0           &36.4            &41.9     \\ \hline
CANF-VC        &58.2           &73.0           &48.8         &116.8        &87.5           &56.3            &60.5     \\ \hline
DCVC           &115.7          &150.8          &106.4        &257.5        &118.6          &129.5           &103.9     \\ \hline
DCVC-TCM       &32.8           &62.1           &29.0         &75.8         &25.4           &23.1            &38.2      \\ \hline
DCVC-HEM       &--0.7          &16.1           &--7.1        &20.9         &--15.6         &--17.2          &--1.6     \\ \hline
DCVC-DC        &--13.9         &--8.8          &--27.7       &--19.1       &--27.9         &--25.9          &--14.4    \\ \hline
DCVC-FM        &--8.8         &--5.0          &--23.3       &\bf{--20.8}       &--18.6         &--20.5 &--7.4    \\ \hline
DCVC-SDD     &--13.7         &--2.3          &--24.9       &--8.4        &--17.5         &--19.7          &--7.1    \\ \hline
Ours           &\bf{--20.3}    &\bf{--10.4}    &\bf{--29.2}  &--15.0  &\bf{--25.9}    &\bf{--29.2}     &\bf{--16.7}\\ 
\bottomrule[1.5pt]
\end{tabular}}
\label{table:ip32_psnr}
\end{table*}
\begin{table*}[!htb]
\caption{BD-rate(\%) comparison in RGB colorspace measured with PSNR. The anchor is DCVC-FM. \textbf{96 frames with intra period is --1.} }
  \centering
\scalebox{1}{
\begin{tabular}{l|c|c|c|c|c|c|c}
\toprule[1.5pt]
               & HEVC Class B  & HEVC Class C  &HEVC Class D &HEVC Class E &HEVC Class RGB &UVG            &MCL-JCV  \\ \hline
DCVC-FM        &0.0            &0.0            &0.0          &\bf{0.0}          &0.0            &0.0             &0.0      \\ \hline
DCVC-DC        &--1.0          &5.3            &2.8          &26.1         &--4.7          &5.5             &--0.1    \\ \hline
DCVC-SDD       &8.1            &23.9           &10.0         &118.7        &20.0           &24.5            &13.7    \\ \hline
Ours           &\bf{--13.9}    &\bf{--8.7}    &\bf{--9.8}  &15.0     &\bf{--9.4}    &\bf{--11.6}     &\bf{--11.0}\\ 
\bottomrule[1.5pt]
\end{tabular}}
\label{table:ip96_psnr}
\end{table*}
\begin{table*}[!htb]
\caption{BD-rate(\%) comparison in RGB colorspace measured with PSNR. The anchor is DCVC-FM. \textbf{All frames with intra period is --1.} }
  \centering
\scalebox{1}{
\begin{tabular}{l|c|c|c|c|c|c|c}
\toprule[1.5pt]
               & HEVC Class B  & HEVC Class C  &HEVC Class D &HEVC Class E &HEVC Class RGB &UVG            &MCL-JCV  \\ \hline
DCVC-FM            &0.0            &0.0            &0.0          &\bf{0.0}          &0.0            &0.0             &0.0      \\ \hline
DCVC-DC        	   &14.9           &27.2           &21.5         &152.1        &21.7           &21.4            &2.1   \\ \hline
DCVC-SDD     	   &128.2          &88.1           &44.2         &Cannot Calculate          &106.6          &60.2            &18.3    \\ \hline
Ours               &\bf{--12.2}    &\bf{--1.7}    &\bf{--5.9}  &30.1  &\bf{--7.0}    &\bf{--11.2}     &\bf{--12.1}\\  

\bottomrule[1.5pt]
\end{tabular}}
\label{table:ip1_psnr}
\end{table*}

\begin{figure*}[t]
  \centering
  \begin{minipage}[c]{0.9\linewidth}
  \centering
  \includegraphics[width=\linewidth]{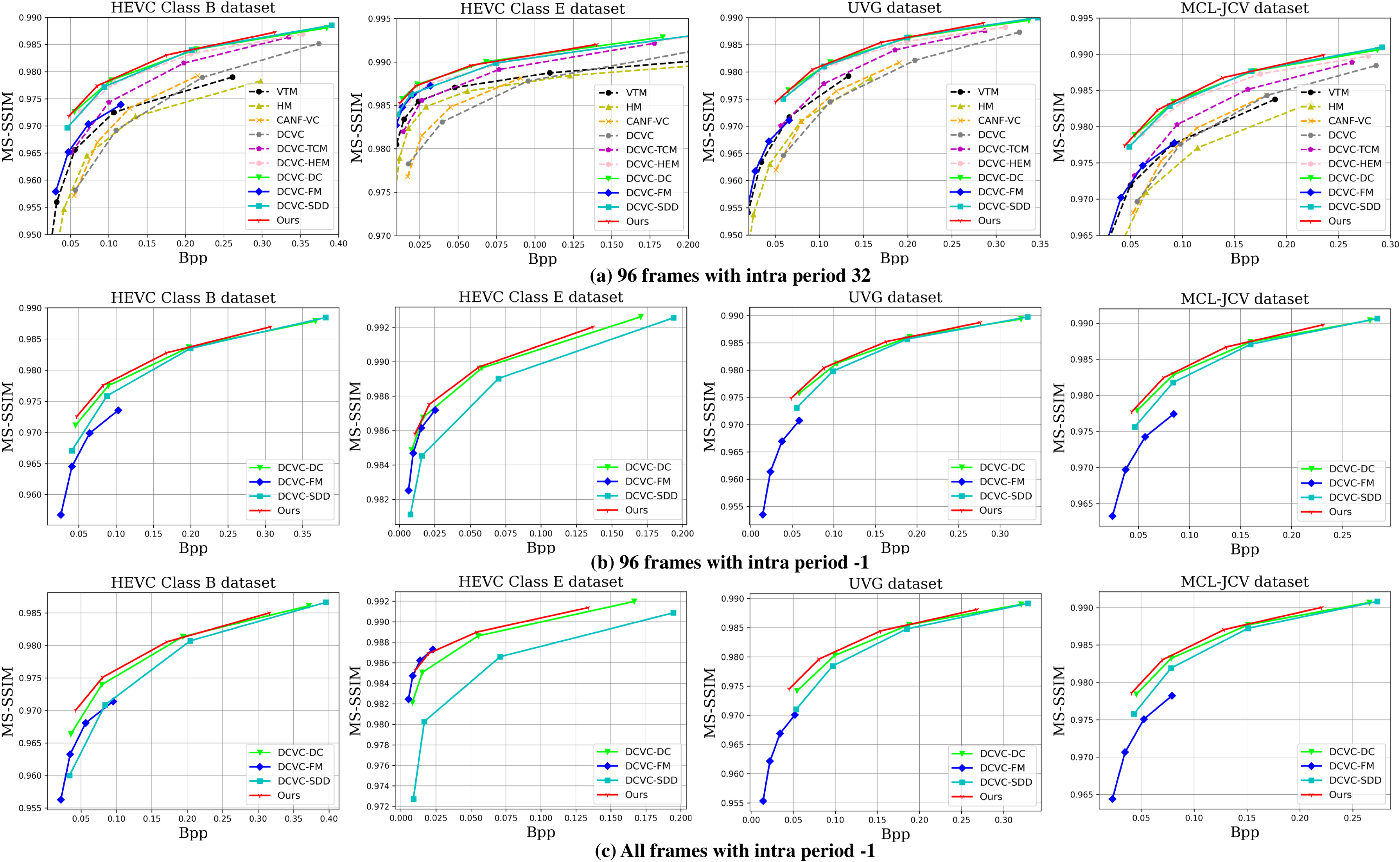}
 \end{minipage}%
    \caption{Rate-distortion curves when the quality is measured by MS-SSIM.}
\label{fig:msssim}
\end{figure*}
\begin{table*}[!htb]
\caption{BD-rate(\%) comparison measured with MS-SSIM. The anchor is VTM. 96 frames with intra period is 32.} 
  \centering
\scalebox{1}{
\begin{threeparttable}
\begin{tabular}{l|c|c|c|c|c|c|c}
\toprule[1.5pt]
               & HEVC Class B  & HEVC Class C  &HEVC Class D &HEVC Class E &HEVC Class RGB &UVG            &MCL-JCV  \\ \hline
VTM            &0.0            &0.0            &0.0          &0.0          &0.0            &0.0             &0.0      \\ \hline
HM             &36.8           &38.7           &34.9         &38.4         &37.3           &37.1            &43.7     \\ \hline
CANF-VC        &25.5           &17.7           &1.5         &114.9        &52.9           &33.1            &11.7     \\ \hline
DCVC           &35.9          &24.9          &2.7        &90.0        &43.7          &11.9           &39.1     \\ \hline
DCVC-TCM       &--20.5         &--21.7           &--36.2         &--20.5         &--21.1           &--6.0            &--18.6      \\ \hline
DCVC-HEM       &--47.4          &--43.3           &--55.5        &--52.4         &--45.8         &--32.7          &--44.0     \\ \hline
DCVC-DC        &--53.0         &--54.6          &--63.4       &--60.7       &--54.4         &--36.7          &--49.1    \\ \hline
DCVC-FM        &--12.5         &--18.0          &--30.6       &--32.6       &--16.6         &--7.3          &--5.0    \\ \hline
DCVC-SDD     &--48.0         &--49.6          &--60.0       &--51.5        &--46.3         &--34.2          &--46.3    \\ \hline
Ours           &\bf{--55.6}    &\bf{--55.4}    &\bf{--64.1}  &\bf{--59.1}  &\bf{--54.3}    &\bf{--39.7}     &\bf{--51.0}\\ 
\bottomrule[1.5pt]
\end{tabular}
  \begin{tablenotes}
   \item \footnotesize \dag Due to the MS-SSIM model of DCVC-FM is not released, when the reconstruction quality is measured by MS-SSIM, we use the PSNR model of DCVC-FM to supplement the result, although this is unfair.
  \end{tablenotes}
\end{threeparttable}}
\label{table:ip32_msssim}
\end{table*}
\begin{table*}[!htb]
\caption{BD-rate(\%) comparison in RGB colorspace measured with MSSSIM. The anchor is DCVC-FM. \textbf{96 frames with intra period is --1.} } 
  \centering
\scalebox{1}{
\begin{tabular}{l|c|c|c|c|c|c|c}
\toprule[1.5pt]
               & HEVC Class B  & HEVC Class C  &HEVC Class D &HEVC Class E &HEVC Class RGB &UVG            &MCL-JCV  \\ \hline
DCVC-FM        &0.0            &0.0            &0.0          &0.0          &0.0            &0.0             &0.0      \\ \hline
DCVC-DC        &--41.1          &--42.0            &--44.9          &--14.9         &--34.3          &--41.8             &--45.5    \\ \hline
DCVC-SDD       &--23.1            &--26.3           &--31.2         &65.7        &0.5           &--33.0            &--25.4    \\ \hline
Ours           &\bf{--48.2}    &\bf{--47.0}    &\bf{--48.1}  &\bf{--21.3}     &\bf{--41.9}    &\bf{--46.7}     &\bf{--50.4}\\ 
\bottomrule[1.5pt]
\end{tabular}}
\label{table:ip96_msssim}
\end{table*}
\begin{table*}[!htb]
\caption{BD-rate(\%) comparison in RGB colorspace measured with MSSSIM. The anchor is DCVC-FM. \textbf{All frames with intra period is --1.} } 
  \centering
\scalebox{1}{
\begin{tabular}{l|c|c|c|c|c|c|c}
\toprule[1.5pt]
               & HEVC Class B  & HEVC Class C  &HEVC Class D &HEVC Class E &HEVC Class RGB &UVG            &MCL-JCV  \\ \hline
DCVC-FM            &0.0            &0.0            &0.0          &\bf{0.0}          &0.0            &0.0             &0.0      \\ \hline
DCVC-DC        	   &--28.6           &--31.7           &--35.9         &63.2        &--15.3           &--34.7            &--43.8   \\ \hline
DCVC-SDD     	   &18.9          &4.9           &--1.8         &388.4          &63.4          &--7.2            &--28.1    \\ \hline
Ours               &\bf{--45.5}    &\bf{--41.9}    &\bf{--44.7}  &7.8  &\bf{--38.4}    &\bf{--49.6}     &\bf{--50.5}\\ 

\bottomrule[1.5pt]
\end{tabular}}
\label{table:ip1_msssim}
\end{table*}
\begin{figure*}[t]
  \centering
  \begin{minipage}[c]{0.9\linewidth}
  \centering
  \includegraphics[width=\linewidth]{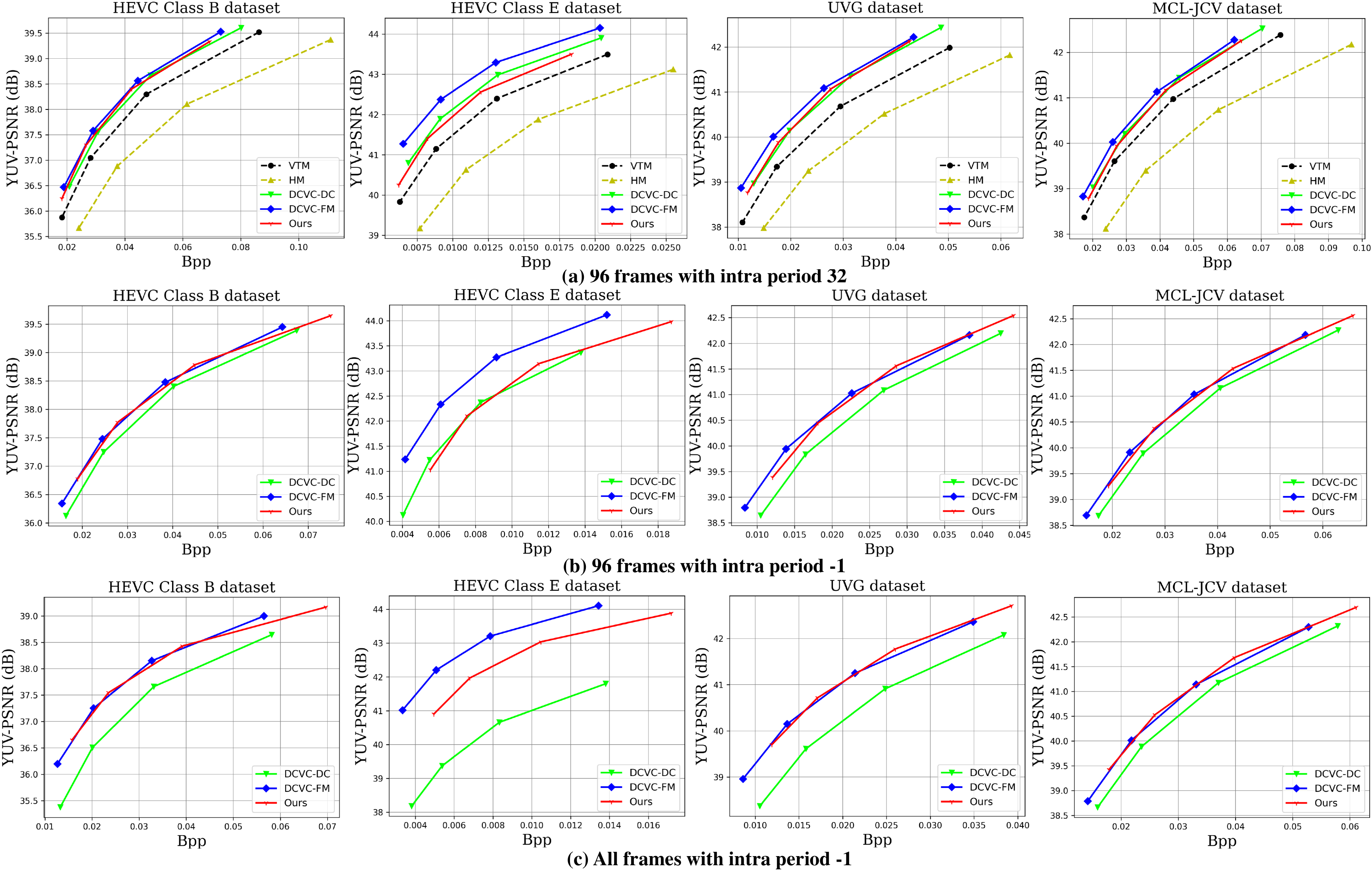}
 \end{minipage}%
    \caption{Rate-distortion curves when the quality is measured by YUV PSNR.}
\label{fig:yuv_psnr}
\end{figure*}
\begin{table*}[htb]
\caption{BD-rate(\%) comparison in YUV colorspace measured with PSNR. The anchor is VTM. 96 frames with intra period is 32. }
  \centering
\scalebox{1}{
\begin{tabular}{l|c|c|c|c|c|c}
\toprule[1.5pt]
               & HEVC Class B  & HEVC Class C  &HEVC Class D &HEVC Class E &UVG            &MCL-JCV  \\ \hline
VTM            &0.0            &0.0            &0.0          &0.0           &0.0                &0.0      \\ \hline
HM             &41.0           &36.1           &31.9         &44.6          &37.7               &43.2    \\ \hline
DCVC-DC        &--11.6         &--13.1         &--28.8       &--18.1                  &--17.2             &--11.0    \\ \hline
DCVC-FM       &\bf{--16.6}         &\bf{--17.7}         &\bf{--33.4}       &\bf{--29.9}                 &\bf{--25.0}             &\bf{--15.6} \\ \hline
Ours           &--14.0    &--13.7    &--29.8  &--13.2   &--18.3     &--9.8\\ 
\bottomrule[1.5pt]
\end{tabular}}
\label{table:ip32_yuv_psnr}
\end{table*}

\begin{table*}[htb]
\caption{BD-rate(\%) comparison in YUV colorspace measured with PSNR. The anchor is DCVC-FM. \textbf{96 frames with intra period is --1.}}
  \centering
\scalebox{1}{
\begin{tabular}{l|c|c|c|c|c|c}
\toprule[1.5pt]
               & HEVC Class B  & HEVC Class C  &HEVC Class D &HEVC Class E &UVG            &MCL-JCV  \\ \hline
DCVC-FM        &\bf{0.0}       &\bf{0.0}       &\bf{0.0}     &\bf{0.0}     &\bf{0.0}       &\bf{0.0}      \\ \hline
DCVC-DC        &10.2           &13.4           &13.9         &35.3         &20.0            &10.5   \\ \hline
Ours           &0.9            &4.9            &6.2          &34.9         &2.9             &0.7\\ 
\bottomrule[1.5pt]
\end{tabular}}
\label{table:ip96_yuv_psnr}
\end{table*}

\begin{table*}[!htb]
\caption{BD-rate(\%) comparison in YUV colorspace measured with PSNR. The anchor is DCVC-FM. \textbf{All frames with intra period is --1.} }
  \centering
\scalebox{1}{
\begin{tabular}{l|c|c|c|c|c|c}
\toprule[1.5pt]
               & HEVC Class B  & HEVC Class C  &HEVC Class D &HEVC Class E &UVG            &MCL-JCV  \\ \hline
DCVC-FM        &\bf{0.0}       &\bf{0.0}       &\bf{0.0}     &\bf{0.0}     &0.0       &0.0      \\ \hline
DCVC-DC        &34.7           &33.4           &30.8         &201.6        &36.3            &11.8   \\ \hline
Ours           &1.2            &4.6            &7.0          &46.7         &\bf{--0.8}    &\bf{--1.8}\\ 
\bottomrule[1.5pt]
\end{tabular}}
\label{table:ip1_yuv_psnr}
\end{table*}
We use the Lagrangian multiplier $\lambda$ to control the rate-distortion (R-D) trade-off. We also add a periodically varying weight $w_t$ for each P-frame before the Lagrangian multiplier $\lambda$ to implement the hierarchical quality~\cite{sheng2024spatial, li2023neural}.  The detailed setting of $\lambda$ and $w_t$ can be found in the Section~\ref{implementation_details}.
\par
When using $L_{t}^{meD}$ as the loss function, we only train the motion estimation module, motion encoder, and motion decoder (Inter). When using $L_{t}^{meRD}$ as the loss function, we add the motion entropy model into the training loop (Inter). When using $L_{t}^{recD}$ as the loss function, we only train the temporal context mining module, contextual encoder, contextual decoder, and frame generator (Rec).  When using $L_{t}^{recRD}$ as the loss function, we add the contextual entropy model into the training loop (Rec). When using $L_{t}^{all}$ or $L_{T}^{all}$ as the loss function, we train all modules.
\par
When the quality of reconstructed RGB videos is measured by MS-SSIM, we use the RGB-MSE model before being trained with $L_T^{all}-repeat-long$ as the pre-trained model. Then, we replace the distortion metric from MSE to 1--MS-SSIM to fine-tune the pre-trained model. The detailed fine-tuning strategy is listed in Table.~\ref{training_stategy_ssim}.\

When encoding YUV420 videos, we use the RGB-MSE model before being trained with $L_T^{all}$ in RGB colorspace as the pre-trained model. Then, we fine-tune the pre-trained model with the loss functions calculated in the YUV colorspace. The detailed fine-tuning strategy is listed in Table.~\ref{training_stategy_yuv}.\par

\section{Experiments}\label{sec:experiments}
\subsection{Experimental Setup}
\subsubsection{Training Data}
We follow most existing learned video coding schemes to use 7-frame videos of Vimeo-90k~\cite{xue2019video} dataset for short-sequence training.  We also follow M-LVC~\cite{lin2020m} to generate some sequences containing 19 frames from raw Vimeo videos for long-sequence training. We randomly crop the original videos into 256$\times$256 patches.

\subsubsection{Testing Data}
We use videos in RGB format and YUV420 format to evaluate the performance of our proposed scheme. When testing YUV420 videos, we use the original YUV420-format UVG dataset~\cite{mercat2020uvg}, MCL-JCV dataset~\cite{wang2016mcl}, and HEVC dataset~\cite{bossen2013common}. 
We feed them into learned video codecs without any change. When testing RGB videos, we convert UVG, MCL-JCV, and HEVC datasets from YUV420 format to RGB format using FFmpeg and then feed them into learned video codecs. In addition, following~\cite{sheng2022temporal,li2023neural}, we also test 6 RGB-format videos from HEVC RGB dataset~\cite{flynn2015overview}.

\subsubsection{Implementation Details}\label{implementation_details}
We implement our proposed methods based on our baseline~\cite{sheng2024spatial}. Following~\cite{sheng2024spatial, li2022hybrid, li2023neural}, we first set 4 basic $\lambda$ values (85, 170, 380, 840) to control the R-D trade-off.  For each $\lambda$, four learnable quantization steps are embedded into the motion encoder, motion decoder, contextual encoder, and contextual decoder. During testing, we interpolate the quantization steps to support variable rates~\cite{sheng2024spatial, li2023neural}.
For hierarchical quality structure, we follow our baseline~\cite{sheng2024spatial} and set the hierarchical weight $w_t$ as (0.5, 1.2, 0.5, 0.9) for  4 consecutive frames. 
The weight $w_t$ of the first p frame is 1.2. We set the number of training frames to 6 or 19 to set the weight $w_t$ of the last training frame as 1.2. For different $w_t$, the parameters of the first convolutional layer of the temporal context mining module are not shared~\cite{sheng2024spatial, li2023neural}.\par
When PSNR is evaluated in RGB colorspace, we feed the original Vimeo training frames in RGB format into our model to obtain reconstructed frames. Then we calculate the distortion (MSE) between input frames and reconstructed frames in RGB format to train our model. 
When the PSNR is evaluated in YUV420 colorspace, we first convert the format of Vimeo training frames from RGB to YUV444 and feed the frames in YUV444 format into our model to obtain reconstructed frames. Then we convert the format of input frames and reconstructed frames from YUV444 to YUV420 and calculate their MSE in YUV420 format to train our model. Although the weight of compound evaluation YUV PSNR is 6:1:1~\cite{ohm2012comparison}, we find that setting the weight of MSE of YUV components to 4:1:1 during training can achieve more stable results. We implement our model with PyTorch. The AdamW~\cite{kingma2014adam} optimizer is used and the batch size is set to 8. 

\subsubsection{Test configurations}
We focus on the low-delay coding scenario in this paper. Following~\cite{li2024neural}, we test 96 frames and all frames for each video sequence. The intra period is set to 32 and --1. When testing RGB videos, we compare with CANF-VC~\cite{ho2022canf}, DCVC~\cite{li2021deep}, DCVC-TCM~\cite{sheng2022temporal},  DCVC-HEM~\cite{li2022hybrid},DCVC-DC~\cite{li2023neural}, DCVC-FM~\cite{li2024neural}, and our baseline---DCVC-SDD~\cite{sheng2024spatial}. When testing YUV420 videos, we only compare with DCVC-DC~\cite{li2023neural} and DCVC-FM~\cite{li2024neural} since they are the only learned video codecs that have released models for the YUV420 colorspace. Since DCVC-FM is a variant bit rate model and its bit range is much wider, we try to align its bit rate range with that of other learned video codecs by adjusting its \emph{q\_index\_i} and \emph{q\_index\_p}. Specifically, for RGB PSNR and MS-SSIM metrics, \emph{q\_index\_i} and \emph{q\_index\_p} are set to 42, 49, 56, and 63. For YUV PSNR, these parameters are configured as 37, 44, 51, and 59.\par

We also compare with traditional video codecs, including the reference software of H.265/HEVC---HM-16.20~\cite{HM} and the reference software of H.266/VVC---VTM-13.2~\cite{VTM}. When testing RGB videos, for HM-16.20, we use \emph{encoder\_lowdelay\_main\_rext} configuration. For VTM-13.2, we use \emph{encoder\_lowdelay\_vtm} configuration. Following~\cite{sheng2022temporal}, we set the internal colorspace as YUV444 for better compression performance. When testing YUV420 videos, for HM-16.20, we use \emph{encoder\_lowdelay\_main} configuration. For VTM-13.2, we use \emph{encoder\_lowdelay\_vtm} configuration. Both HM-16.20 and VTM-13.2 are under the low-delay B reference structure.
The detailed commands for HM-16.20 and VTM-13.2 are shown as follows.
\begin{itemize}
    \item -c $\{\emph{config file name}\}$  \mbox{-}\mbox{-}InputFile=$\{\emph{input file name}\}$  \mbox{-}\mbox{-}FrameRate=$\{\emph{frame rate}\}$ \mbox{-}\mbox{-}InputBitDepth=8 \mbox{-}\mbox{-}IntraPeriod=$\{\emph{intra period}\}$ \mbox{-}\mbox{-}DecodingRefreshType=2 
 \mbox{-}\mbox{-}FramesToBeEncoded=$\{\emph{frames}\}$ \mbox{-}\mbox{-}QP=$\{\emph{qp}\}$ \mbox{-}\mbox{-}SourceWidth=$\{\emph{width}\}$  \mbox{-}\mbox{-}SourceHeight=$\{\emph{height}\}$ 
 \mbox{-}\mbox{-}Level=6.2 \mbox{-}\mbox{-}BitstreamFile=$\{\emph{bitstream file name}\}$
\mbox{-}\mbox{-}InputChromaFormat=$\{\emph{color format}\}$
\end{itemize}
\par

\begin{figure*}[htb]
  \centering
   \includegraphics[width=0.8\linewidth]{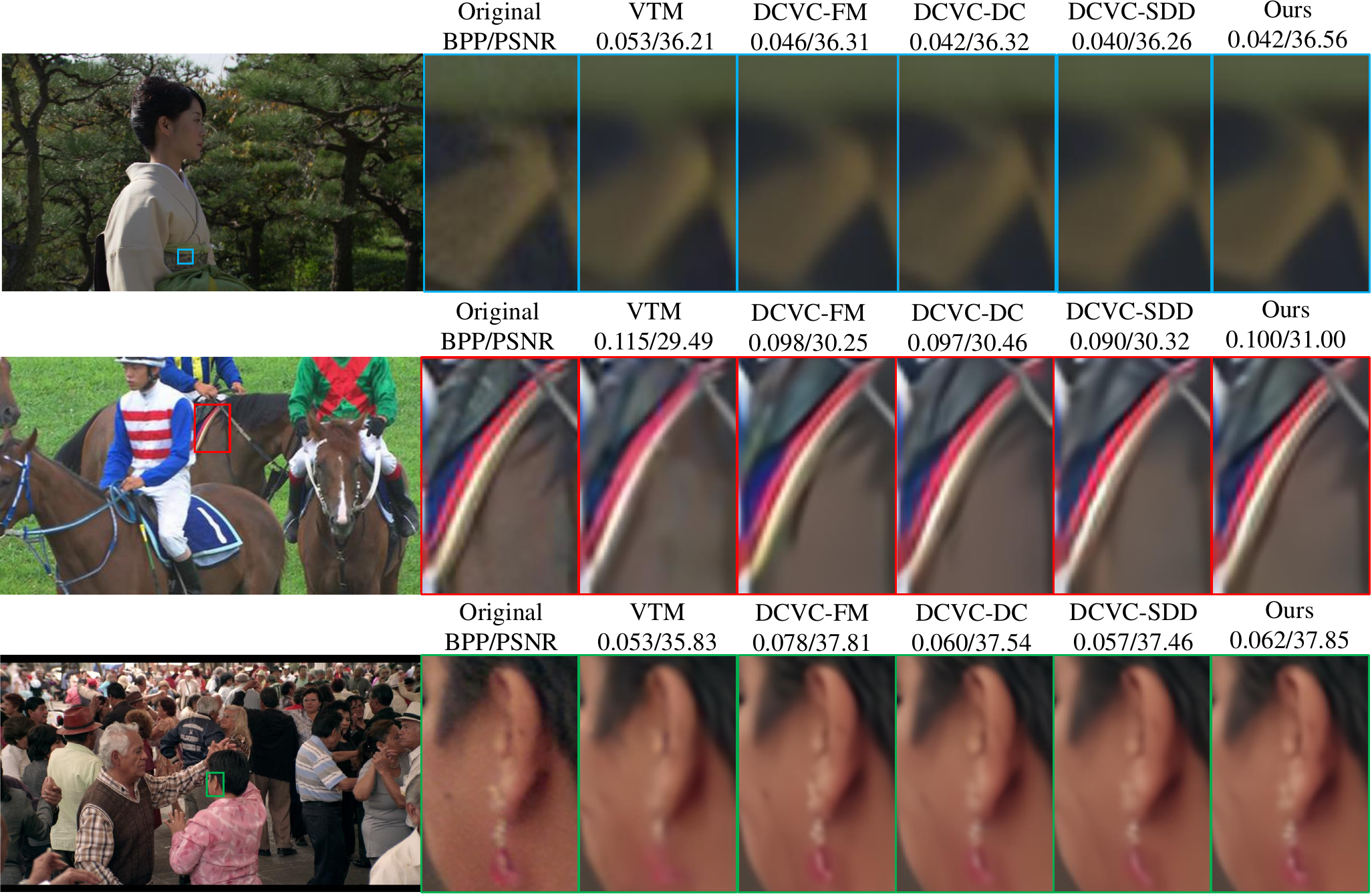}
      \caption{Subjective quality comparison on the 3rd frame of the HEVC Class B {\em Kimono1} sequence, the 7th frame of HEVC Class D {\em RaceHorses} sequence, and the 5th frame of MCL-JCV {\em videoSRC14} sequence.}
   \label{fig:subjective}
\end{figure*}

\subsubsection{Evaluation Metrics}
When testing RGB videos, we use RGB PSNR and MS-SSIM to measure the distortion between reconstructed videos and original frames. When testing YUV420 videos, following \cite{li2023neural, li2024neural}, we use compound YUV PSNR  as the distortion metric. The weight of YUV components is 6:1:1~\cite{ohm2012comparison}. We use bits per pixel (bpp) to measure the average number of bits for encoding each pixel in each frame.\par

\subsection{Experimental Results}
\subsubsection{Objective Comparison Results for RGB Colorspace}
When testing RGB videos and using RGB PSNR to measure the distortion, we present the rate-distortion curves under different testing conditions in Fig.~\ref{fig:psnr_results} and list the detailed BD-rate values in Table~\ref{table:ip32_psnr}, Table~\ref{table:ip96_psnr}, and Table~\ref{table:ip1_psnr}. The results show that  our proposed scheme outperforms our baseline DCVC-SDD~\cite{sheng2024spatial} by a large margin and even
achieves a better compression performance than previous SOTA schemes--DCVC-DC and DCVC-FM. For example, when the intra period is --1, our scheme achieves --12.2\% bitrate saving against DCVC-FM over the HEVC Class B dataset. When testing RGB videos and using MS-SSIM to measure the distortion, we present the rate-distortion curves in Fig.~\ref{fig:msssim} and list the BD-rate comparison in Table~\ref{table:ip32_msssim}, Table~\ref{table:ip96_msssim}, and Table~\ref{table:ip1_msssim}. The comparison results also verify the larger performance improvement of our scheme. It also outperforms our baseline DCVC-SDD~\cite{sheng2024spatial}, DCVC-DC~\cite{li2023neural}, and DCVC-FM~\cite{li2024neural}. Although our scheme achieves better performance on most testing datasets, we find it perform worse than DCVC-FM on HEVC Class E dataset.  This may be because the training dataset we used lacks videos like Class E where the foreground has tiny movements and the background does not move. DCVC-FM introduced a new training dataset for 32-frame cascaded fine-tuning but this dataset is not released. Our collected videos for 19-frame fine-tuning have similar characteristics to the original Vimeo 90k. We will try our best to collect more videos to solve this problem in the future.

\subsubsection{Objective Comparison Results for YUV420 Colorspace}
When compressing YUV420 videos and using YUV PSNR to measure the distortion, we illustrate the rate-distortion curves in Fig.~\ref{fig:yuv_psnr} and list the corresponding BD-rate comparison results in Table~\ref{table:ip32_yuv_psnr}, Table~\ref{table:ip96_yuv_psnr}, and Table~\ref{table:ip1_yuv_psnr}. The results show that the compression performance of our scheme is better than that of DCVC-DC on most testing datasets but is worse than that of DCVC-FM. This may be because DCVC-FM has a much longer YUV fine-tuning cycle than our YUV model and has a more complex training strategy. We will further explore to improve our YUV compression performance in the future.

\subsubsection{Subjective Comparison Results}
We illustrate the reconstructed frames of HEVC Class B {\em Kimono1}, HEVC Class D {\em RaceHorses} sequence, and MCL-JCV {\em videoSRC14} sequence in Fig.~\ref{fig:subjective}.
By comparing the reconstructed frames of VTM, DCVC-FM~\cite{li2024neural}, DCVC-DC~\cite{li2023neural}, DCVC-SDD~\cite{sheng2024spatial}, and our scheme, we can observe that our scheme can reconstruct clearer textures with similar bitrate cost. For example, the pattern of the belt in our reconstructed \emph{Kimono1} sequence and the edge of the saddle of the horse in our reconstructed \emph{RaceHorses} sequence are sharper. The earring of the woman in our reconstructed \emph{videoSRC14} sequence retains more details.\par
\begin{table}[t]
 \centering
 \caption{Runtime and model complexity comparison for a 1080p video frame.}
\scalebox{1}{
\begin{tabular}{c|c|c|c|c}
\toprule[1.5pt]
Schemes  & Enc Time & Dec Time  & MACs/pixel & Model Size          \\ \hline
DCVC  & 14.96 s& 44.01 s & 1159.34K & 6.53M \\ \hline
DCVC-TCM  & 0.81 s& 0.48 s & 1609.75K & 10.55M\\ \hline
DCVC-HEM  & 0.75 s& 0.26 s & 1791.64K& 17.52M \\ \hline
DCVC-DC     & 0.82 s& 0.64 s & 1397.90K&18.45M \\ \hline
DCVC-FM     & 0.74 s& 0.53 s &1180.77K &17.02M \\ \hline
DCVC-SDD      & 0.94 s& 0.74 s & 1849.06K& 18.74M \\ \hline
Ours     & 1.11 s& 0.85 s  & 1963.56K& 19.28M               \\
\bottomrule[1.5pt]
\end{tabular}}
\label{time}
\end{table}

\subsection{Running Time and Model Complexity}
Following the setting of previous learned video coding schemes~\cite{sheng2022temporal,sheng2024spatial,li2023neural,li2022hybrid}, we include the time for model inference, entropy modeling, entropy coding, and data transfer between CPU and GPU when calculating the encoding and decoding time. Table.~\ref{time} lists the detailed encoding and decoding time and computational complexities for a 1920$\times$1080 video frame of different learned video codecs. We run all these codecs on a NVIDIA 3090 GPU. The comparison results show that our proposed technologies lead to only 0.17s encoding time and 0.11s decoding time increase compared with our baseline DCVC-SDD~\cite{sheng2024spatial}. In addition, the computational complexity and the number of trainable parameters of our codec are 1963.56K MACs/pixel and 19.28M, respectively, which is not a significant increase compared to our baseline.
\begin{figure*}[t]
  \centering
   \includegraphics[width=0.85\linewidth]{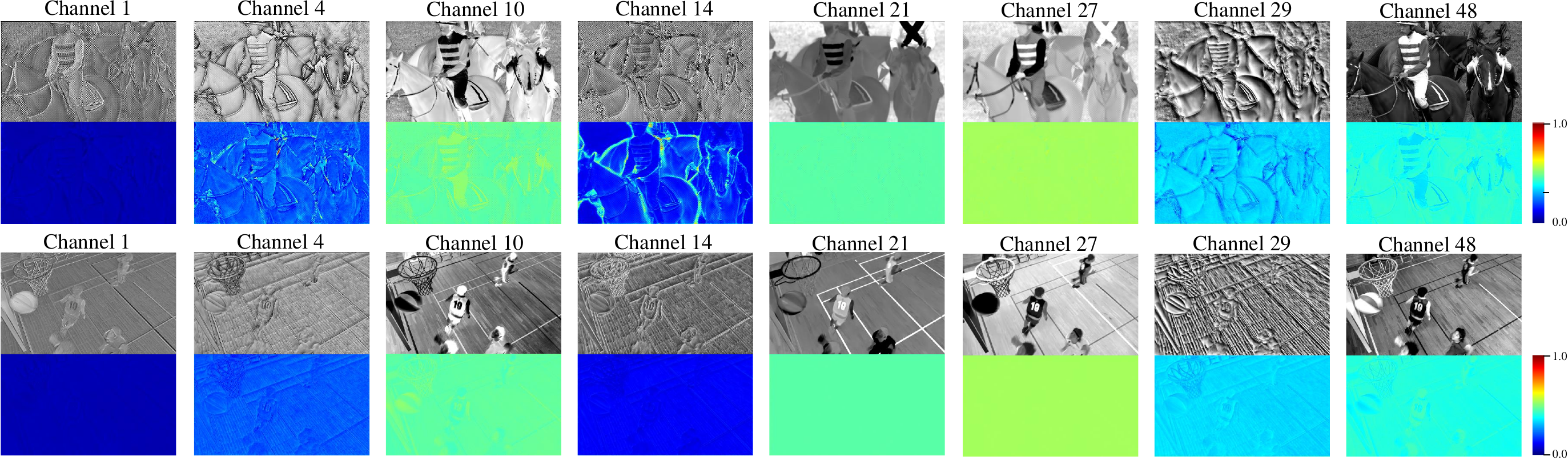}
      \caption{Visualization of temporal contexts of different channels and their corresponding confidence maps.}
   \label{fig:PQA_visualization}
\end{figure*}
\begin{figure*}[t]
  \centering
   \includegraphics[width=0.85\linewidth]{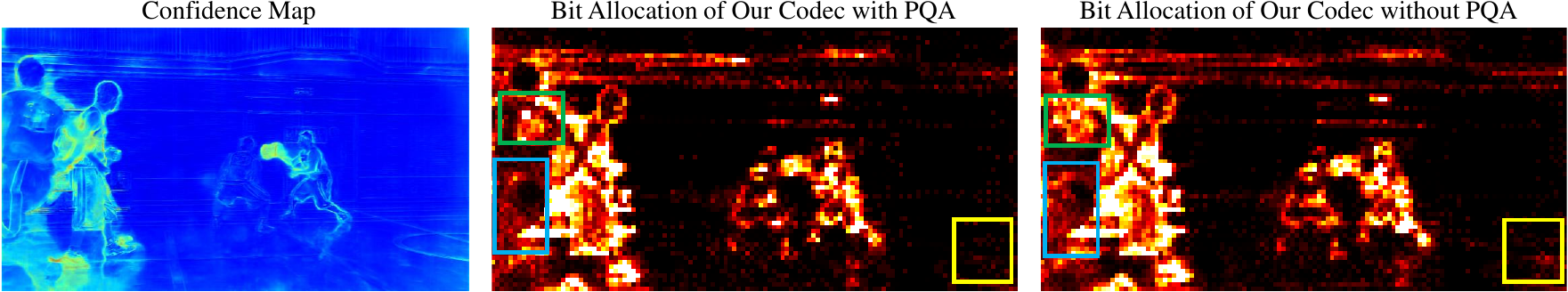}
      \caption{Comparison of bit allocation maps of our codec with and without the PQA module.}
   \label{fig:PQA_bit_allocation_visualization}
\end{figure*}
 \begin{table}[t]
\caption{Effectiveness of Proposed Technologies.}
\centering
\scalebox{1}{
\begin{tabular}{c|c|c|c|c|c}
\toprule[1.5pt]
Model Index &PQA & RQA &Repeat &Long & BD-Rate(\%)\\ \hline
$M_1$&\XSolidBrush    &\XSolidBrush &\XSolidBrush &\XSolidBrush & 0.0  \\ \hline
$M_2$&\Checkmark    &\XSolidBrush &\XSolidBrush &\XSolidBrush   &--2.1 \\ \hline
$M_3$&\Checkmark    &\Checkmark &\XSolidBrush &\XSolidBrush     & --4.6    \\\hline
$M_4$&\XSolidBrush    &\XSolidBrush &\Checkmark &\XSolidBrush     & --1.0   \\\hline
$M_5$&\XSolidBrush    &\XSolidBrush &\Checkmark &\Checkmark     & --3.7    \\\hline
$M_6$&\Checkmark    &\Checkmark &\Checkmark &\XSolidBrush       &--5.4    \\\hline
$M_7$&\Checkmark    &\Checkmark &\XSolidBrush &\Checkmark & --7.5  \\ \hline
$M_8$&\Checkmark    &\Checkmark &\Checkmark &\Checkmark   & --8.9    \\
\bottomrule[1.5pt]
\end{tabular}
}
\label{effectiveness}
\end{table}
\subsection{Ablation Studies}
\subsubsection{Effectiveness of Proposed technologies}
In this paper, based on our baseline DCVC-SDD~\cite{sheng2024spatial}, we propose a prediction quality adaptation module, a reference quality adaptation module, and a training strategy that combines repeating compressing and long-sequence cascaded training. To verify the effectiveness of these proposed technologies, we conduct an ablation study on the HEVC dataset by progressively enabling these technologies, as presented in Table~\ref{effectiveness}.
When calculating BD-rate, we regard our baseline ($M_1$) as the anchor. The testing intra period is set to 32 and the distortion is measured by RGB-PSNR.  Comparing Model $M_1$, $M_2$, and $M_3$, we find that enabling our proposed prediction quality adaptation (PQA) module can bring 2.1\% BD-Rate reduction. Enabling the proposed reference quality adaptation (RQA) brings an additional 2.5\% performance improvement. To verify the effectiveness of our proposed new training strategies, based on Model $M_3$, we add the repeat-compressing strategy and long-sequence training strategy respectively. Comparing Model $M_3$ and $M_6$, applying the repeat-compressing strategy only brings 0.8\% performance gain. Comparing Model $M_3$ and $M_7$, applying the long-sequence training strategy can bring a 2.9\% performance gain. If we apply the two strategies simultaneously, comparing Model $M_3$ and $M_8$, a higher performance gain (4.3\%) can be achieved. For a supplement, we also apply the training strategies to our baseline. Comparing Model $M_1$, $M_4$, and $M_5$, we observe that the training strategies also bring benefits to the baseline. The comparison between $M_5$ and $M_8$ further verifies the effectiveness of our proposed PQA and RQA modules. 

\begin{figure}[t]
  \centering
   \includegraphics[width=0.5\linewidth]{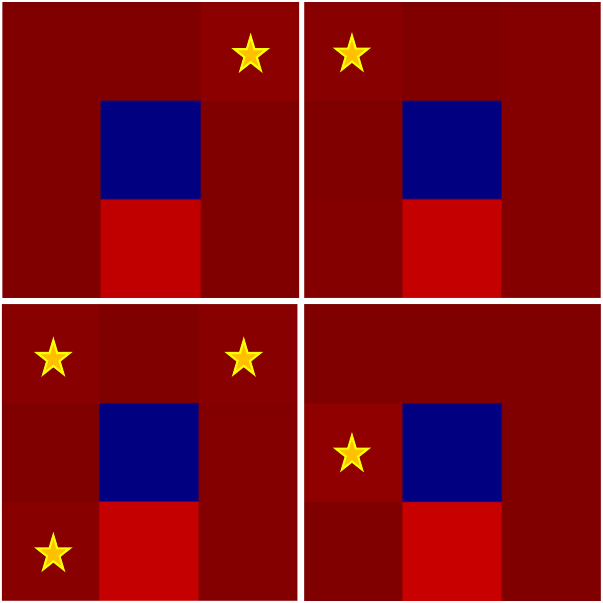}
      \caption{Visualization of the dynamic filters of different frames learned by the RQA module. Asterisks are used to indicate the filter areas where there are significant differences.}
   \label{fig:RQA_visualization}
\end{figure}
\begin{figure*}[t]
  \centering
  \begin{minipage}[c]{\linewidth}
  \centering
  \includegraphics[width=\linewidth]{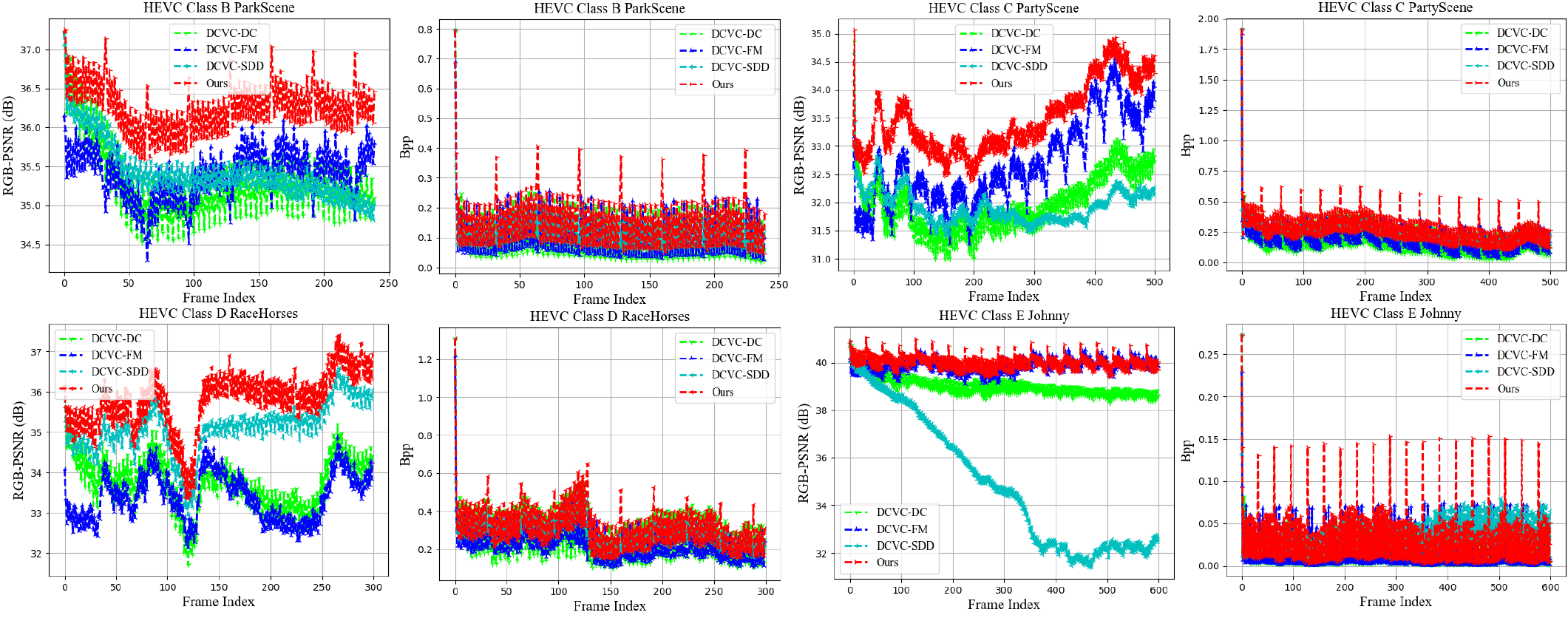}
 \end{minipage}%
    \caption{Reconstructed quality and bit rate comparison across frames under similar bit rate points.}
  \label{fig:RQA_ablation}
\end{figure*}

\subsubsection{Analysis of Prediction Quality Adaptation Module}\label{ablation:PQA}
To explore why our proposed PQA module can bring performance improvement, we visualize the temporal contexts generated by the model with and without the PQA module. As shown in Fig.~\ref{fig:PQA_visualization}, we can observe that spatial-wise prediction difference exists for the temporal context of a certain channel. For example, the confidence map of the temporal context of the $14^{th}$ channel has larger confidence values at the edges of objects, which indicates that the temporal context of this channel mainly provides prediction for high-frequency object edges. We also analyze the relationship between the confidence map and bit allocation maps. As shown in Fig.~\ref{fig:PQA_bit_allocation_visualization}, we can observe that the confidence maps can help the codec focus on the regions with higher confidence values and benefit the bit reduction of the corresponding regions in the frame. In addition, comparing different channels of the temporal contexts, we find that channel-wise prediction difference also exists for the temporal contexts. For the channels with higher prediction qualities, such as the $10^{th}$,  $21^{st}$, $27^{th}$, and $48^{th}$ channels, the values of their confidence maps are larger, which means they provide the main temporal information. If they cannot provide enough temporal information for some regions, other channels provide additional supplementary information. For most other channels, their temporal prediction has lower qualities and only provide limited temporal information, resulting the smaller values for their confidence maps. The analysis verifies that our proposed PQA module can provide explicit spatial and channel-wise discrimination for temporal contexts. In addition, this analysis also indicates that the learned video codecs do not need such a large number of temporal context channels. This provides a direction for light-weighting of learned video codecs such as channel pruning.\par

\subsubsection{Analysis of Reference Quality Adaptation Module}\label{ablation:RQA}
To analyze why our proposed RQA module can bring performance gain, we first visualize the dynamic filters learned by the RQA module of different frames. As shown in Fig.~\ref{fig:RQA_visualization}, we can observe that our proposed RQA module can learn different filters from different reference frames. As described in Section~\ref{sec:RQA}, these dynamic filters can help the transform networks of our codec adapt to different reference qualities and adjust the implicit quantization, making it easier for our codec to achieve the target reconstruction quality controlled by $\lambda$ in the loss function. To further verify the effectiveness of our proposed RQA module, we compare the reconstructed frame qualities and bit rate across frames of our codec with other codecs. As illustrated in Fig.~\ref{fig:RQA_ablation}, we take the sequences in HEVC Class B, C, D, and E datasets as examples. Although DCVC-SDD and DCVC-DC have added a periodically varying weight before $\lambda$ in the loss function to adapt to different reference qualities, their reconstructed frame qualities still gradually decrease as the reference qualities decrease. In addition, compared with DCVC-FM which uses 32 frames for cascaded training, our codec can achieve a comparable reduction in error propagation with only 19 frames. This is mainly because our proposed RQA module can help our codec adapt to different reference qualities, making it easier to achieve the target reconstructed frame quality. These analyses verify the effectiveness of our RQA module.

\section{Conclusion}\label{sec:conclusion}
In this paper, we first propose a confidence-based prediction quality adaptation module to adapt to different prediction qualities. With this module, our codec can learn spatial and channel-wise confidence maps to adaptively decide which spatial or channel location of predictions to use. Then, we further propose a reference quality adaptation module and an associated repeat-long training strategy to provide dynamic spatially variant filters for diverse reference qualities. With this module, our codec can adapt to different reference qualities, making it easier to achieve the target reconstruction quality and reduce reconstruction error propagation. Experimental results show that our codec can achieve better compression performance.
\bibliographystyle{ieeetr}
\bibliography{ref}
\begin{IEEEbiography}[{\includegraphics[width=1in,height=1.25in,clip,keepaspectratio]{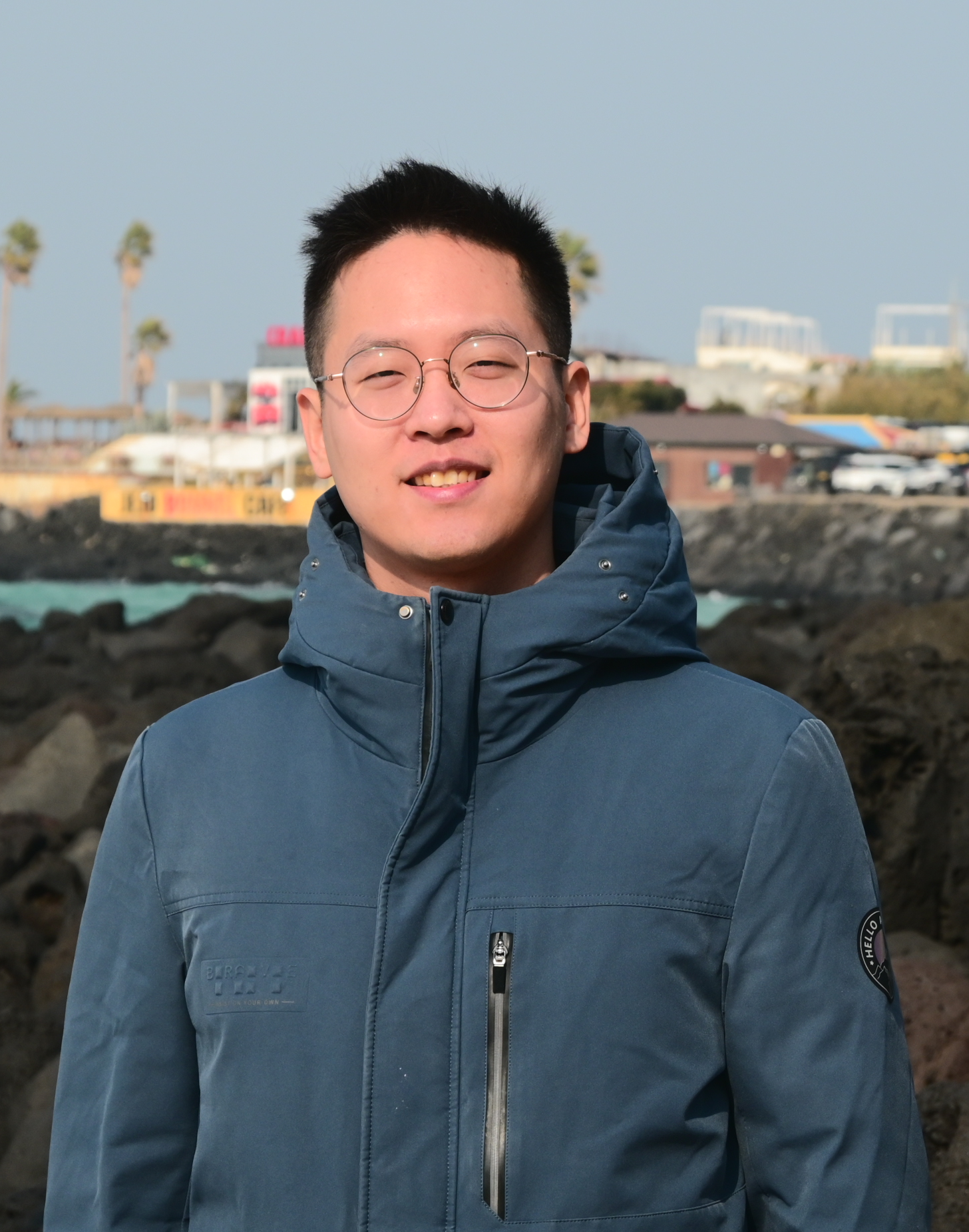}}]{Xihua Sheng} (M'24) received the B.S. degree in automation from Northeastern University, Shenyang, China, in 2019, and the Ph.D. degree in electronic engineering from University of Science and Technology of China (USTC), Hefei, Anhui, China, in 2024. 
He is currently a Postdoctoral Fellow in computer science from City University of Hong Kong. 
His research interests include image/video/point cloud coding, signal processing, and machine learning.
\end{IEEEbiography}

\begin{IEEEbiography}[{\includegraphics[width=1in,height=1.25in,clip,keepaspectratio]{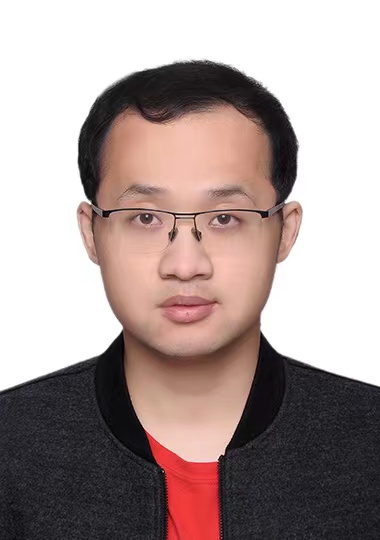}}] {Li Li} (M'17--SM'25) received the B.S. and Ph.D. degrees in electronic engineering from University of Science and Technology of China (USTC), Hefei, Anhui, China, in 2011 and 2016, respectively.
He was a visiting assistant professor in University of Missouri-Kansas City from 2016 to 2020.
He joined the department of electronic engineering and information science of USTC as a research fellow in 2020 and became a professor in 2022.

His research interests include image/video/point cloud coding and processing.
He has authored or co-authored more than 80 papers in international journals and conferences. 
He has more than 20 granted patents. 
He has several technique proposals adopted by standardization groups.
He received the Multimedia Rising Star 2023.
He received the Best 10\% Paper Award at the 2016 IEEE Visual Communications and Image Processing (VCIP) and the 2019 IEEE International Conference on Image Processing (ICIP).
He serves as an associate editor for \textsc{IEEE Transactions on Circuits and Systems for Video Technology} and \textsc{IEEE Transactions on Multimedia}. 
\end{IEEEbiography}

\begin{IEEEbiography}[{\includegraphics[width=1in,height=1.25in,clip,keepaspectratio]{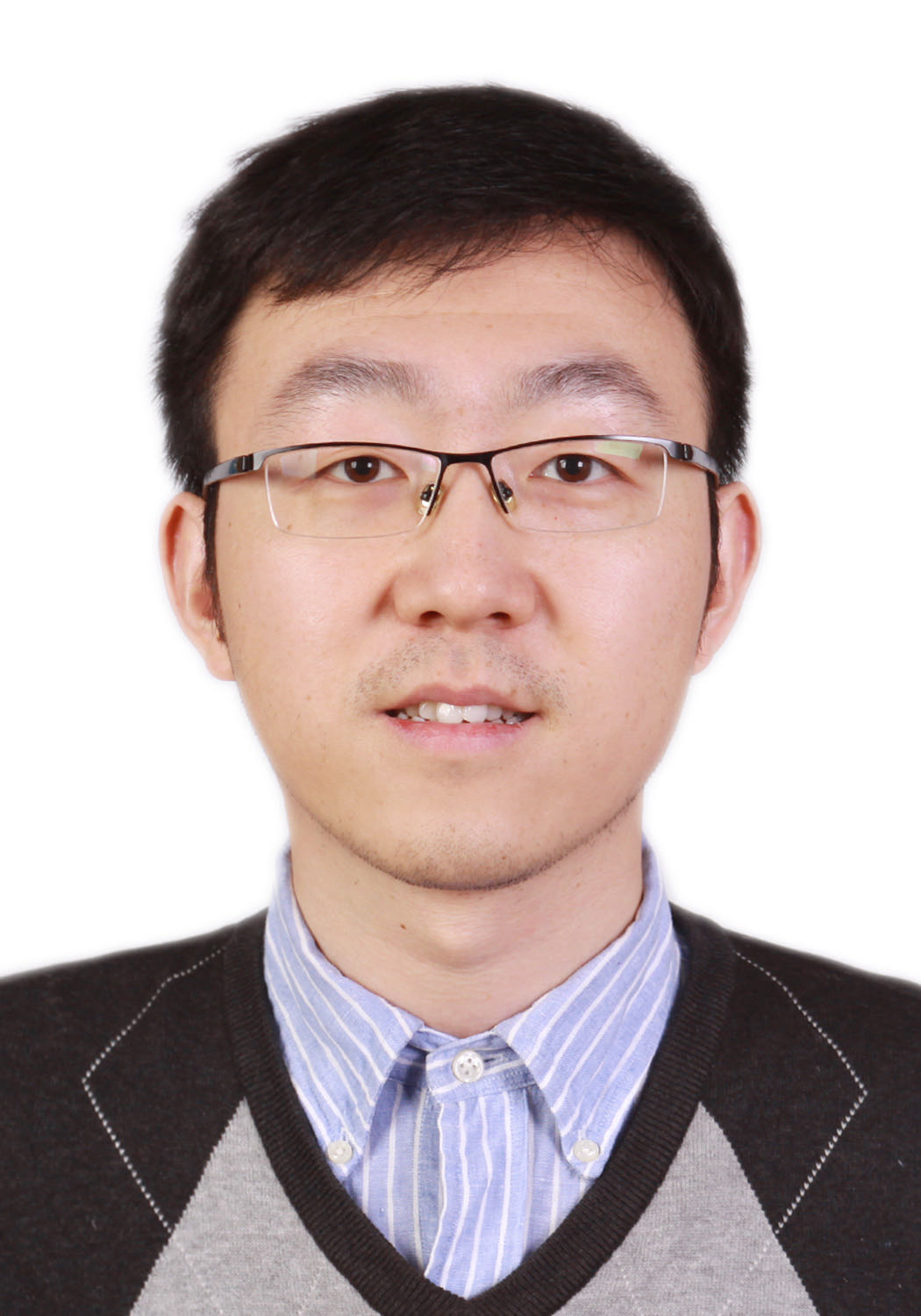}}]{Dong Liu}
(M'13--SM'19) received the B.S. and Ph.D. degrees in electrical engineering from the University of Science and Technology of China (USTC), Hefei, China, in 2004 and 2009, respectively. He was a Member of Research Staff with Nokia Research Center, Beijing, China, from 2009 to 2012. He has been a faculty member at USTC since 2012 and currently holds the position of full professor.

His research interests include image and video processing, coding, and analysis. He has authored or co-authored more than 200 papers in international journals and conferences, which were cited more than 20,000 times according to Google Scholar. He has more than 30 granted patents. He has several technique proposals adopted by standardization groups.

He received 2009 IEEE TCSVT Best Paper Award, VCIP 2016 Best 10\% Paper Award, and ISCAS 2022 Grand Challenge Top Creativity Paper Award. He and his students were winners of several technical challenges held in ICIP 2024, ISCAS 2023, ICCV 2019, etc. He is a Senior Member of CCF and CSIG, an elected member of IVMSP-TC of IEEE SP Society, an elected member of MSA-TC of IEEE CAS Society, and an elected member of Multimedia TC of CSIG. He serves or had served as the Chair of IEEE 1857.11 Standard Working Subgroup (also known as Future Video Coding Study Group), an Associate Editor for \textsc{IEEE Transactions on Image Processing}, a Guest Editor for \textsc{IEEE Transactions on Circuits and Systems for Video Technology}, an Organizing Committee Member for ChinaMM 2024, VCIP 2022, ICME 2021, etc.
\end{IEEEbiography}

\begin{IEEEbiography}[{\includegraphics[width=1in,height=1.25in,clip,keepaspectratio]{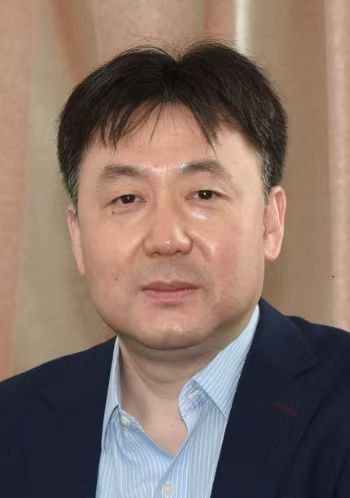}}]{Houqiang Li}
(F'21) is a Professor with the Department of Electronic Engineering and Information Science at the University of Science and Technology of China. His research interests include multimedia search, image/video analysis, video coding, and communication. He has authored and co-authored over 200 papers in journals and conferences. He is the winner of the National Science Funds (NSFC) for Distinguished Young Scientists, the Distinguished Professor of the Changjiang Scholars Program of China, and the Leading Scientist of the Ten Thousand Talent Program of China. He served as an Associate Editor of the IEEE Transactions on Circuits and Systems for Video Technology from 2010 to 2013. He served as the TPC Co-Chair of VCIP 2010, and he served as the General Co-Chair of ICME 2021. He is the recipient of the National Technological Invention Award of China (second class) in 2019 and the recipient of the National Natural Science Award of China (second class) in 2015. He was the recipient of the Best Paper Award for VCIP 2012, the recipient of the Best Paper Award for ICIMCS 2012, and the recipient of the Best Paper Award for ACM MUM in 2011.

Houqiang received the B.S., M. Eng., and Ph.D. degrees in electronic engineering from the University of Science and Technology of China, Hefei, China in 1992, 1997, and 2000, respectively. He was elected as a Fellow of IEEE (2021).
\end{IEEEbiography}

\end{document}